\pdfoutput=1   
\documentclass[preprint,12pt]{elsarticle}

\usepackage{graphicx}
\usepackage{amsmath,amssymb,amsfonts}
\usepackage{bm}
\usepackage{multirow}
\usepackage{booktabs}
\usepackage{tabularx}
\usepackage{enumitem}
\usepackage{xcolor}
\usepackage{url}       
\usepackage[ruled,linesnumbered,noend]{algorithm2e}
\usepackage{nicefrac}       
\usepackage{microtype}      
\usepackage{hyperref}       

\hypersetup{
  pdftitle={Bayesian optimization and topographic exploration of drag-reducing dimples for aerodynamic surfaces},
  pdfauthor={Sangjoon Lee, M. Erden Yildizdag, Haris M. Sheikh},
  pdfsubject={Turbulent drag reduction by Bayesian-optimized dimpled surfaces (LES)},
  pdfkeywords={Bayesian optimization; large eddy simulation; dimple; drag reduction; passive flow control; aerodynamic design}
}

\DeclareMathOperator*{\argmax}{arg\,max}
\DeclareMathAlphabet{\mathbbold}{U}{bbold}{m}{n}
\graphicspath{ {./figures/} }

\biboptions{sort&compress}

\begin{document}

\begin{frontmatter}

\title{Bayesian optimization and topographic exploration of \\ drag-reducing dimples for aerodynamic surfaces}

\author[ntu]{Sangjoon Lee}
\author[itu]{M. Erden Yildizdag}
\author[soton]{Haris M. Sheikh\corref{cor1}}
\ead{h.m.sheikh@soton.ac.uk}
\cortext[cor1]{Corresponding author}

\affiliation[ntu]{organization={School of Mechanical and Aerospace Engineering, Nanyang Technological University}, city={Singapore}, postcode={639798}, country={Singapore}}
\affiliation[itu]{organization={Faculty of Naval Architecture and Ocean Engineering, Istanbul Technical University}, city={Istanbul}, postcode={34469}, country={T\"{u}rkiye}}
\affiliation[soton]{organization={Department of Aeronautical and Astronautical Engineering, University of Southampton}, city={Southampton}, postcode={SO17 1BJ}, country={United Kingdom}}

\begin{abstract}
Dimples offer a promising route to reducing drag on aerodynamic surfaces. However, whether such shallow concavities yield a net benefit depends sensitively on their topography, which demands systematic mapping and exploration of their comprehensive design space. In this study, dimple design is examined as a mixed-variable optimization over four design variables: dimple type, depth, in-plane scale, and streamwise stretch. The design space is explored using MixMOBO, a Bayesian optimizer, coupled with immersed-boundary large eddy simulations of channel flows at a constant flow rate corresponding to a flat channel at a friction Reynolds number of $180$. The optimal solution, a relatively deep, fully packed, streamwise-elongated diamond dimple, attains a $13.2\%$ drag reduction, notably above previously reported values. A Gaussian process metamodel sensitivity analysis identifies dimple topology as the dominant factor, with coverage and elongation acting mainly through interactions, and depth itself carrying no universal sign. A near-wall flow analysis links the leading designs to fully attached, groove-like flow, whereas poorer designs tend to produce local flow separation that incurs adverse form drag. From these findings, key design insights for drag-reducing dimples are provided.
\end{abstract}


\end{frontmatter}
\section{Introduction}
\label{sec1}

Manipulating near-wall flow interactions using structured surface topographies is a pursuit with a pragmatic history in aerodynamic design optimization. A familiar example dates back to the mid-nineteenth century, when golfers empirically observed that nicked and dented balls outperformed smooth ones; this motivated the deliberate use of patterns and later dimpled designs to increase flight distance~\citep{Mehta1985}. Seminal measurements of golf ball lift and drag coefficients and trajectories by \citet{Bearman1976} showed that dimples promote an earlier laminar-to-turbulent transition of the ball surface, delay flow separation, and thereby reduce the pressure (or form) drag that dominates resistance in that configuration. Subsequent work on spherical bodies has further documented how dimple depth, layout and Reynolds number alter the drag coefficient~\citep{Aoyama1994,Penner2003,Choi2006,Smith2010,Aoki2012,Spalensky2015,Abbas2019}.

Dimples, or shallow concavities, have been adopted as passive flow control elements across a broad range of engineering settings. They are implemented onto aircraft wings and airfoils to delay flow separation~\citep{Beves2017,Abdullah2020,Subramanian2022,Xie2022}, and onto downforce surfaces in motorsport to raise the lift-to-drag ratio~\citep{Allarton2020}. In turbomachinery, they provide boundary layer control on low-pressure turbine blades~\citep{Casey2004}. In convective cooling systems, they are considered to enhance convective heat transfer in heat sinks~\citep{Leontiev2017, Isaev2020}. For drag-reducing purposes, many of these applications exploit dimple-induced separation control on curved or bluff bodies, where pressure (form) drag dominates.

In wall-bounded flows of engineering relevance, such as flows in pipelines~\citep{TabkhPaz2018} or over rocket nose cones~\citep{Poirier2024} with substantively thick viscous boundary layers, the total drag originates not only from pressure effects but also from substantial skin friction contributions. Reducing this skin friction drag on aircraft and other vehicle surfaces is a long-standing aerospace objective, since even a modest but robust reduction can translate into appreciable fuel and emission savings. Surface indentations can modify both mechanisms by altering near-wall vorticity, flow separation, and Reynolds stress production. Due to such complexities, whether dimples on a flat or weakly curved surface can yield a net drag reduction remains an open question, with the outcome depending sensitively on the detailed dimple shape and configuration~\citep{Lienhart2008,Gattere2022}.

In fully developed turbulent channel flow, dimple patterns have also been investigated to reduce turbulent drag.~\citet{Tay2015} measured pressure drops in channel experiments and argued that streamwise vorticity introduced by dimples can produce spanwise motion near the wall that stabilizes the near-wall cycle and reduces skin friction, whereas increasing dimple depth promotes flow separation, leading to the growth of form drag. Therefore, the net drag change is a competition between these opposing effects across various design factors; even within the same dimple topography (e.g., spherical dimples) contrasting results can be reported.

Determining whether dimples universally yield drag reduction, however, remains highly subtle. In turbulent boundary layer measurements, shallow spherical dimples evaluated by~\citet{vannesselrooij2016} achieved a drag reduction of up to $4\%$ compared to a flat surface baseline, with a trend anticipating further improvements at higher Reynolds numbers (i.e., higher flow speeds). By contrast, in direct numerical simulation (DNS) conducted by~\citet{Ng2020} for several dimple topographic profiles at a fixed friction Reynolds number of $Re_\tau = 180$, the surface configured with spherical (n.b., termed circular in their paper) dimples was the worst case, resulting in a total drag increase of $6.4\%$, whereas diamond-shaped dimples produced the largest total drag reduction of $7.4\%$. This contrast clearly indicates that the drag reduction performance of shallow dimples is not a simple function of a single parameter but rather results from a complex interplay between geometric and fluid-dynamic features. More recently, \citet{Cetinkaya2025} experimentally showed that the effectiveness of dimpled surfaces in high-Reynolds-number flows is highly sensitive to geometric and flow parameters. Their measurements demonstrated that the drag-reducing performance of dimpled surfaces depends strongly on parameters such as the dimple coverage ratio and the boundary layer thickness, while also highlighting the conflicting conclusions reported throughout the existing literature regarding the underlying flow mechanisms.

Consequently, identifying effective dimple topographies should be framed as a high-dimensional, nonlinear design optimization problem, rather than an effort to establish a universal consensus on whether dimples reduce or increase drag. In other words, the challenge is to isolate drag-reducing dimple topographies from drag-increasing ones while concurrently mapping how correlated design parameters drive performance. The high sensitivity of this design landscape is well-documented, with minor variations in dimple geometry or flow conditions capable of completely reversing the net aerodynamic benefit~\citep{tay2016triangular,spalart2019shallow,Li2020}. Thus, an optimal dimple topography cannot be fully defined without resolving the coupled effects of planform, depth, scale, and spatial arrangement. Moreover, real-world deployment must ultimately navigate long-standing durability and contamination constraints~\citep{spalart2011industry}; these factors, however, lie outside the scope of the present study.

While prior studies, such as~\citet{Ng2020}, provided valuable insights into isolated dimple topographies, this work explores performance across an expansive, multi-parameter design space that systematically expands these baselines. Bypassing an exhaustive search over highly combinatorial topography candidates is necessary due to the large expense of wall-resolved turbulent channel simulations. To address this challenge, we implement a comprehensive pipeline integrating parametric surface generation, wall-resolved large eddy simulations (LES), and adaptive sampling applied to the topographic design of dimples. This strategy leverages Bayesian optimization (BO) coupled with computational fluid dynamics (CFD), which has emerged as a highly sample-efficient framework for navigating expensive, noisy design objectives~\citep{nabae2021bowave, morita2022bocfd, shende2022gebo, Lee2024}.

Throughout, we adopt the channel as the flow setting for this design search. Under viscous scaling, the near-wall region of a channel reproduces the inner layer of an aerodynamic boundary layer, where the near-wall cycle is largely autonomous from the outer flow~\citep{Jimenez1999}; dimples accordingly act within that region. The opposing plane wall fixes the outer length scale, so the flow remains streamwise homogeneous and every design is compared at a common reference condition, which isolates the effect of surface topography from the streamwise growth that a developing boundary layer would introduce.

In this study, we make the following contributions:
\begin{itemize}[nosep,label=$\circ$]
\item The definition and systematic mapping of an expanded, multi-topography dimple design space that encompasses and extends verified literature baselines.
\item The identification of an optimal dimple for drag reduction under a consistent flow rate condition, achieved via a sample-efficient LES+BO framework.
\item The induction of empirical design insights derived from a comprehensive design sensitivity analysis with Bayesian meta-modeling.
\end{itemize}

The remainder of this manuscript is organized as follows. Section~\ref{sec2} outlines the methodology, describing the constitution of the dimpled surface design space, the numerical simulation framework along with its baseline validation, and the adaptive optimization strategy. Section~\ref{sec3} presents the optimization outcomes, beginning with an overview of the design optimization run that identified the optimal drag-reducing configuration; this is followed by a sensitivity analysis to rank geometric influences, a near-wall flow dynamics analysis, and the synthesis of dimple topography design insights. Section~\ref{sec4} provides a broader discussion of these findings, evaluating the interpretation of design parameter interactions and addressing the outlook of this study. Finally, Section~\ref{sec5} summarizes the primary conclusions and outlines future work.

\section{Methodology}
\label{sec2}

\subsection{Dimple Design Space and Optimization Problem}\label{sec2.1}

\begin{figure}[t]
    \centering
    \includegraphics[width=1\linewidth]{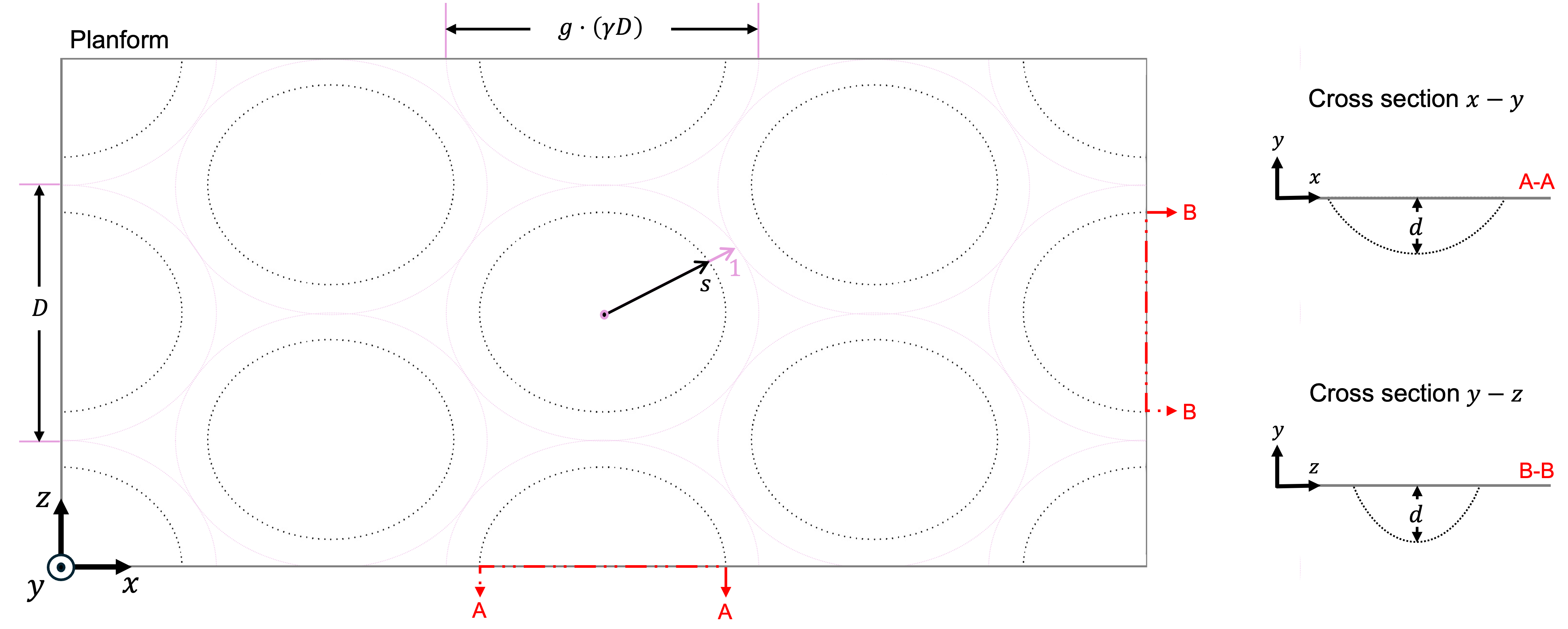}
    \caption{Schematic of the dimpled surface. A repeating dimple unit (black dotted line) is arranged in a staggered manner; $d$, $s$, and $g$ determine the dimple depth, in-plane planform scale with respect to the fully packed arrangement (pink dotted line, $s\rightarrow1$), and streamwise-to-spanwise stretch ratio, respectively. The $s$ arrow is measured against the pink reference arrow labeled $1$, which marks the fully packed extent ($s=1$).}
    \label{fig:surface_schema}
\end{figure}

To configure the dimpled surface, which consists of a repeating array of identical dimples, the topographic concavity profile of each repeat unit is formalized as a systematic design space. This space is parameterized by a mixed variable set: one categorical topography label~$\tau \in \mathcal{T}$ and three continuous geometric controls---dimple depth~$d \in \mathcal{D}$, in-plane planform scale~$s \in \mathcal{S}$, and streamwise-to-spanwise stretch~$g \in \mathcal{G}$. Although $d$, $s$, and $g$ are fundamentally continuous, each is sampled on a prescribed discrete level set so that the BO surrogate can treat depth, scale, and stretch as ordered variables while topology remains an unordered categorical factor. Together, these four input parameters define the non-flat wall surfaces passed to the LES solver to establish the no-slip wall geometry. The spatial arrangement of this configuration is illustrated schematically in Figure~\ref{fig:surface_schema}, adhering to a coordinate convention where $x$, $y$, and $z$ denote the streamwise, wall-normal, and spanwise directions, respectively.

The categorical topographies $\mathcal{T} = \left\{\tau \mid \tau = \mathbbold{0}, \mathbbold{1}, \dots, \mathbbold{6}\right\}$ unify and extend the baseline configurations established in the literature, as indexed in Table~\ref{tab:topography}. This set includes the spherical profile (n.b. termed circular in the notation of~\citet{Ng2020}), downstream- and upstream-oriented teardrop profiles, and the diamond profile. For additional design diversity, the isosceles triangle with a linear depth profile considered by~\citet{tay2016triangular} and its streamwise mirror are included. To isolate depth-profile effects at a fixed circular planform, we introduce a tapered-cylinder profile ($\tau = \mathbbold{0}$) whose planform matches the spherical one ($\tau = \mathbbold{1}$), but features a trapezoidal wall-normal section with a tapered bottom adjusted to match the volume of the spherical cap. It is noted that streamwise stretch of the spherical dimple directly reproduces the elliptical case of~\citet{Ng2020}, eliminating the need for a separate elliptical label.

\begin{table}[t]
\caption{Baseline dimple topographies with their planform and representative cross sections.}
\label{tab:topography}
\begin{tabularx}{\linewidth}{cXccc}
\hline\hline
$\bm{\tau}$ & \textbf{Dimple Type} & \textbf{Planform} & \textbf{Cross Section} $\bm{x-y}$ & \textbf{Cross Section} $\bm{y-z}$ \\
\hline
$\mathbbold{0}$ & Tapered cylinder &
  \parbox[c][1.5cm][c]{2.45cm}{\centering\includegraphics[width=\linewidth]{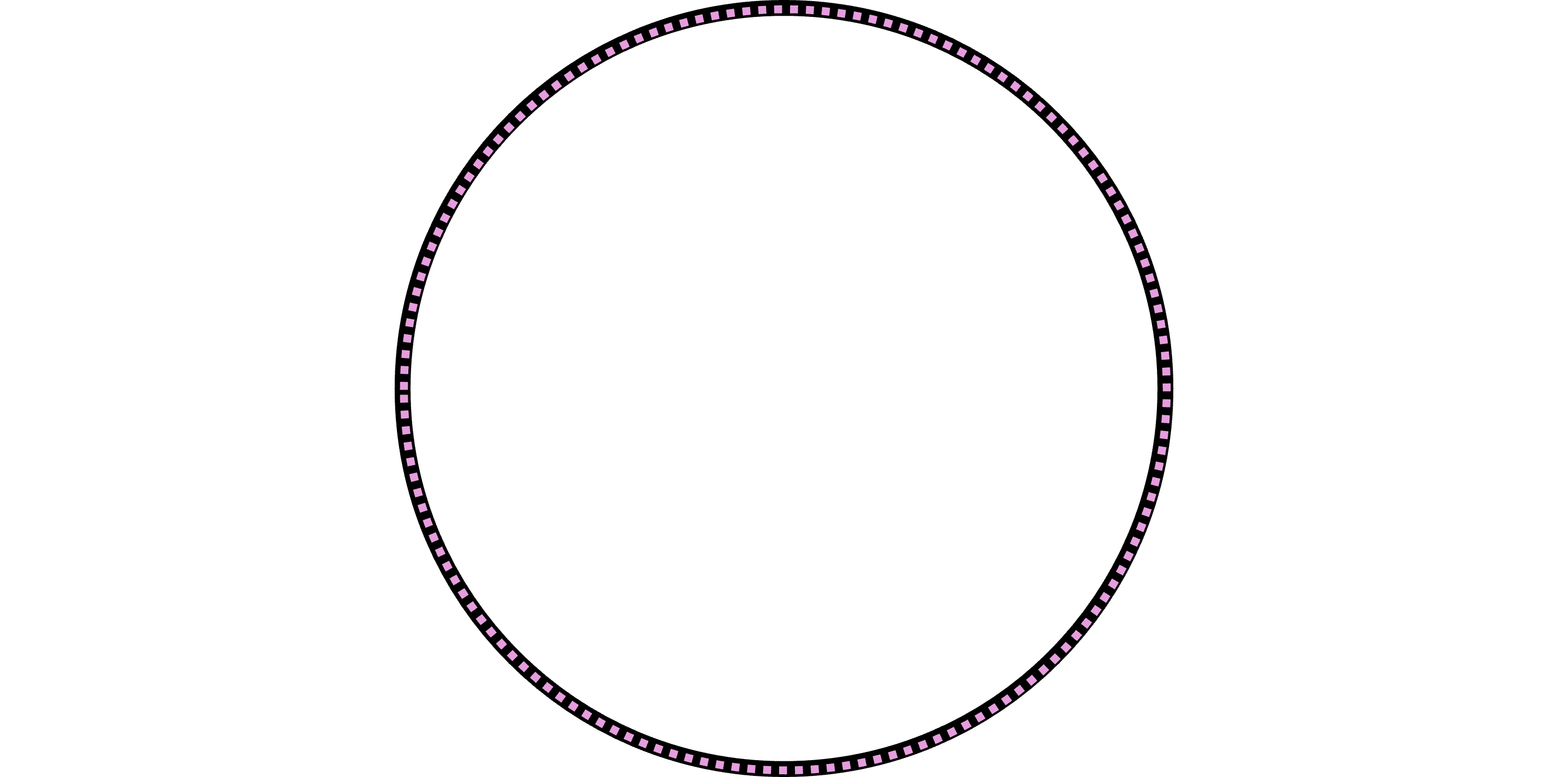}} &
  \parbox[c][1.5cm][c]{2.45cm}{\centering\includegraphics[width=\linewidth]{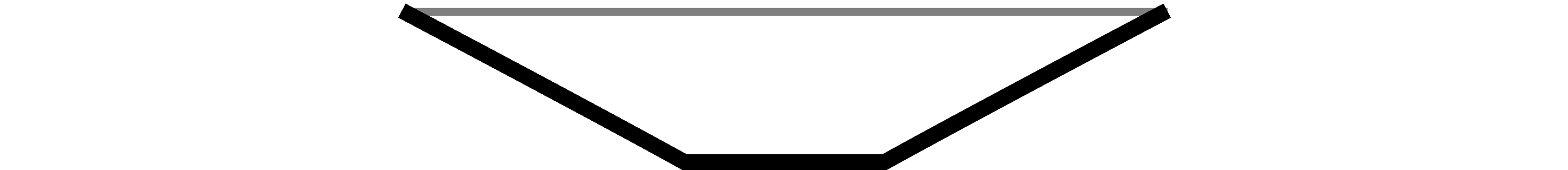}} &
  \parbox[c][1.5cm][c]{2.45cm}{\centering\includegraphics[width=\linewidth]{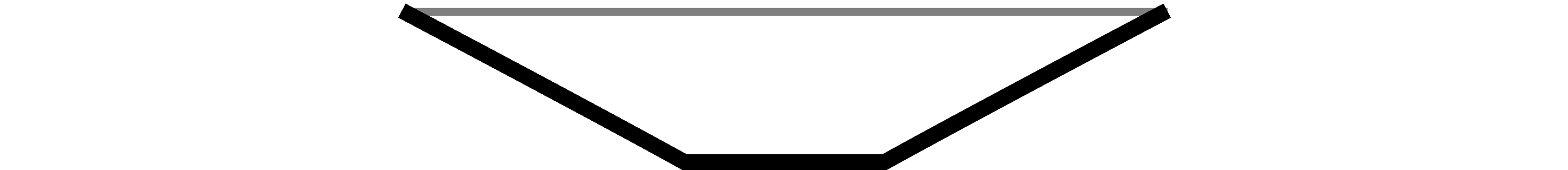}} \\
$\mathbbold{1}$ & Spherical/Elliptical &
  \parbox[c][1.5cm][c]{2.45cm}{\centering\includegraphics[width=\linewidth]{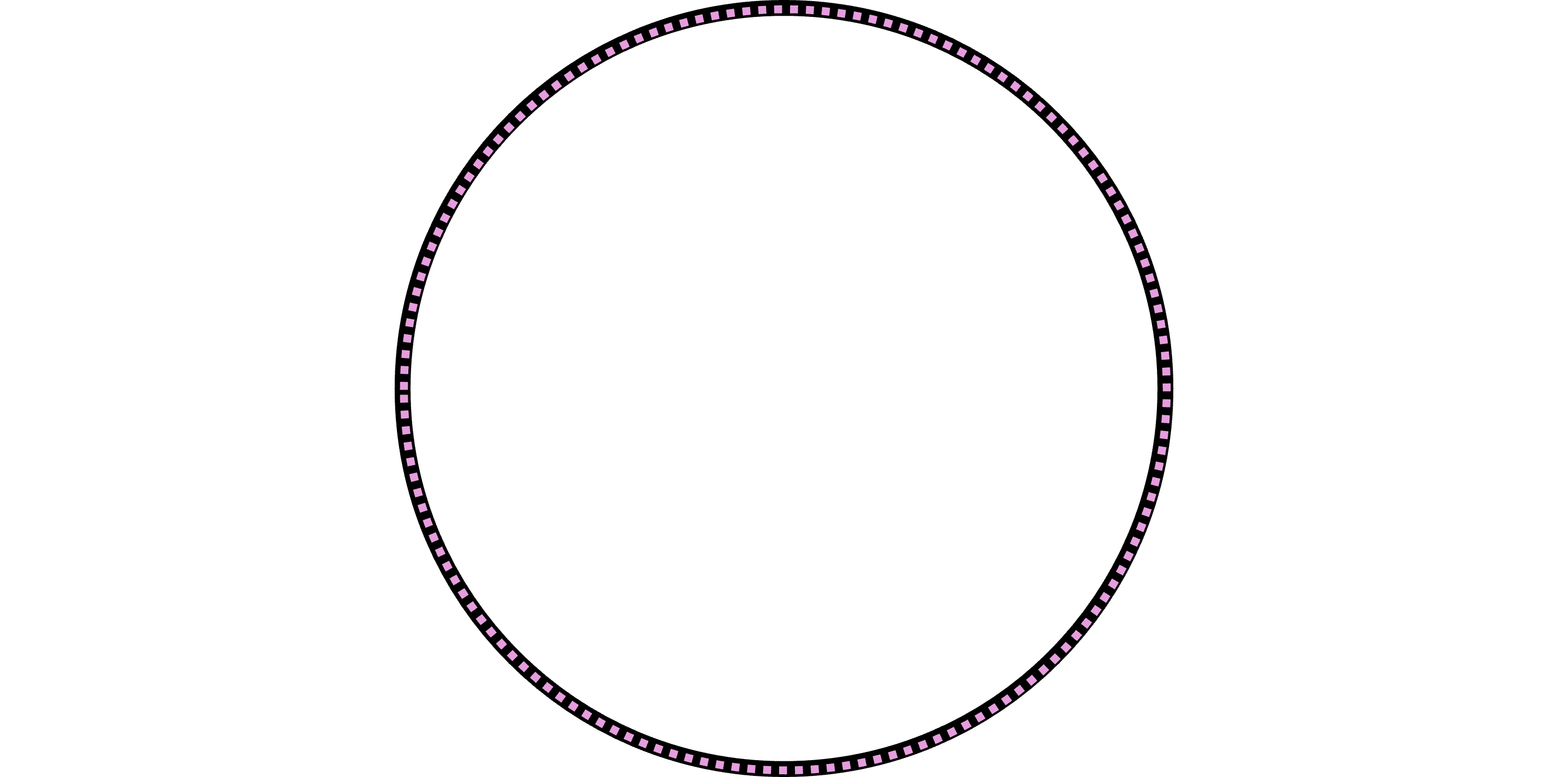}} &
  \parbox[c][1.5cm][c]{2.45cm}{\centering\includegraphics[width=\linewidth]{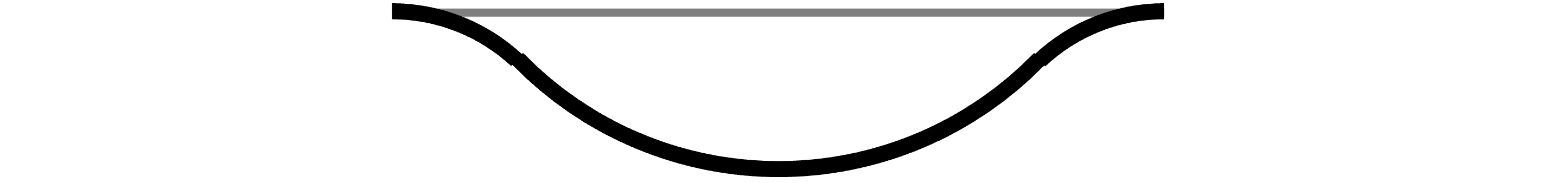}} &
  \parbox[c][1.5cm][c]{2.45cm}{\centering\includegraphics[width=\linewidth]{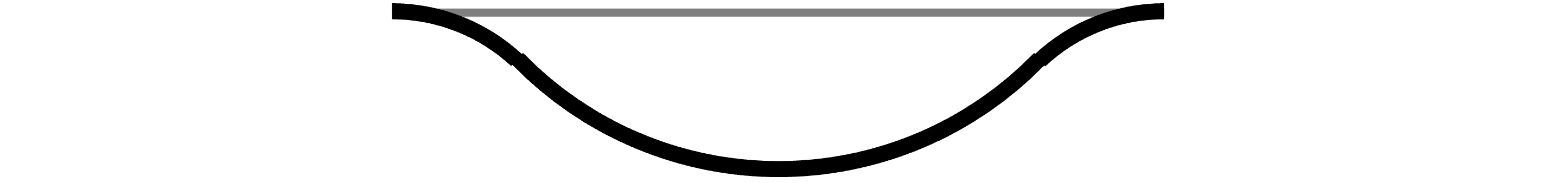}} \\
$\mathbbold{2}$ & Teardrop-down &
  \parbox[c][1.5cm][c]{2.45cm}{\centering\includegraphics[width=\linewidth]{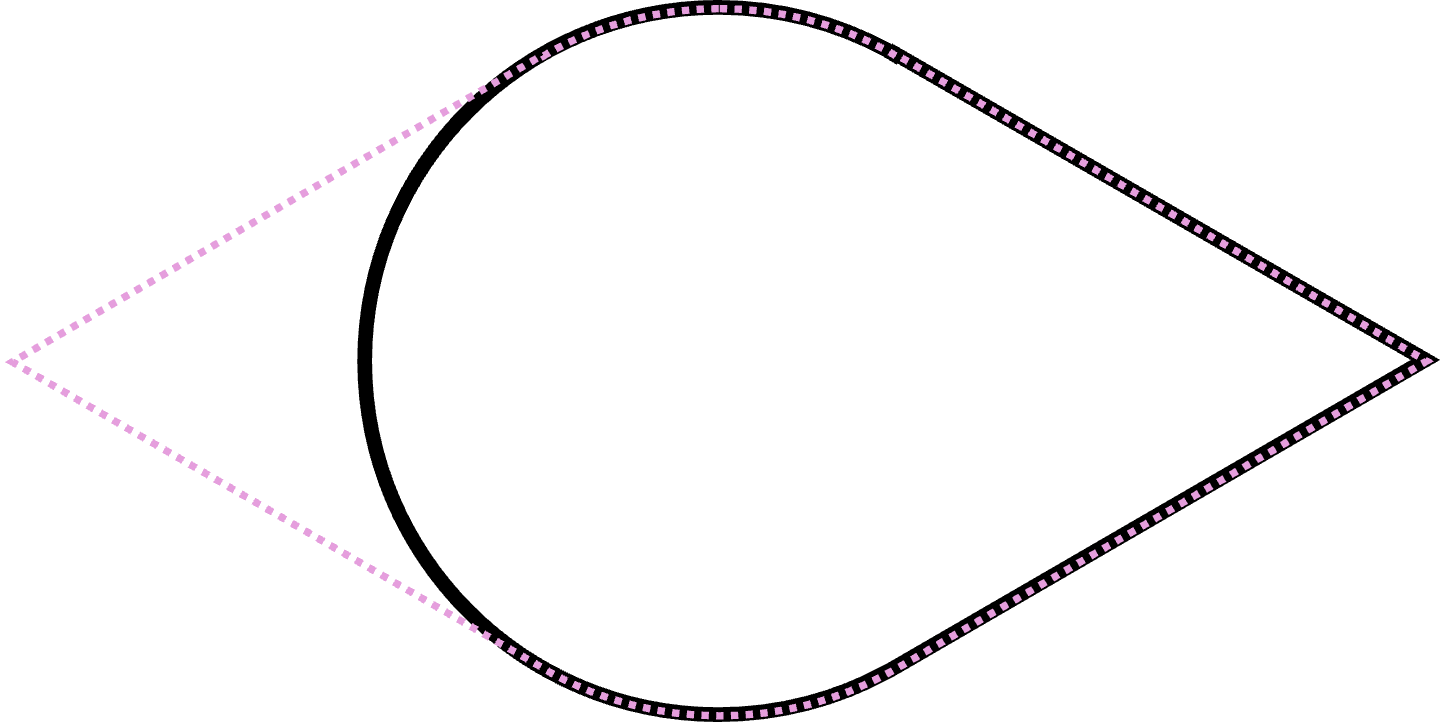}} &
  \parbox[c][1.5cm][c]{2.45cm}{\centering\includegraphics[width=\linewidth]{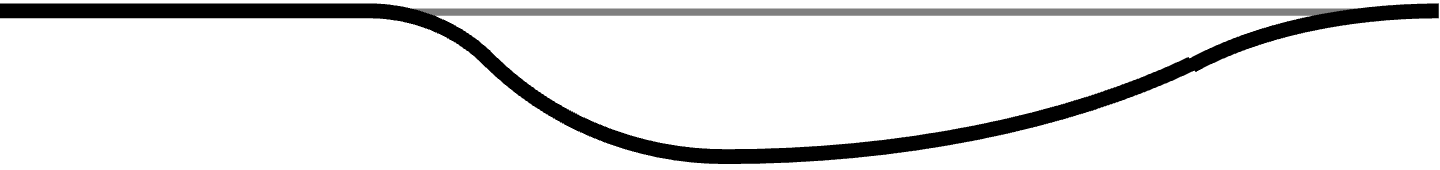}} &
  \parbox[c][1.5cm][c]{2.45cm}{\centering\includegraphics[width=\linewidth]{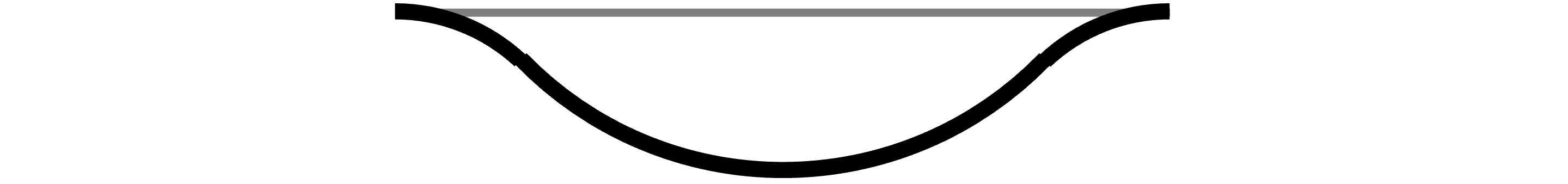}} \\
$\mathbbold{3}$ & Teardrop-up &
  \parbox[c][1.5cm][c]{2.45cm}{\centering\includegraphics[width=\linewidth]{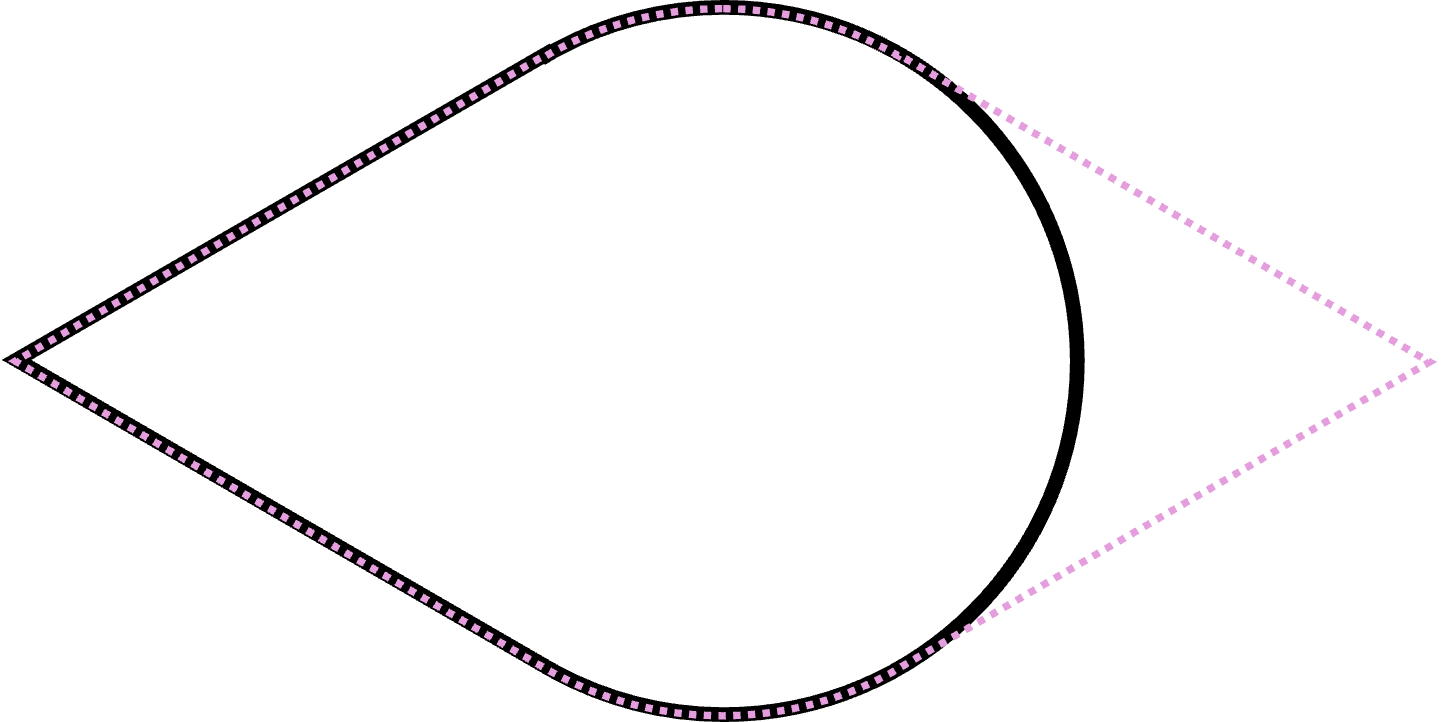}} &
  \parbox[c][1.5cm][c]{2.45cm}{\centering\includegraphics[width=\linewidth]{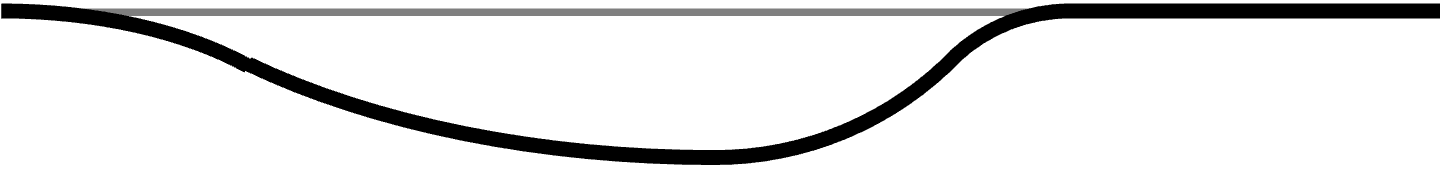}} &
  \parbox[c][1.5cm][c]{2.45cm}{\centering\includegraphics[width=\linewidth]{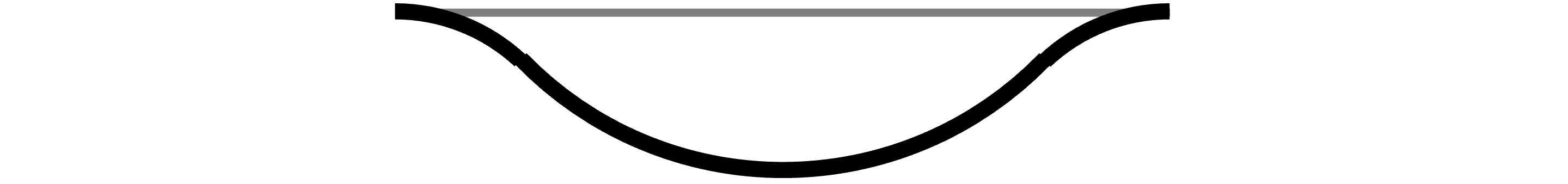}} \\
$\mathbbold{4}$ & Diamond &
  \parbox[c][1.5cm][c]{2.45cm}{\centering\includegraphics[width=\linewidth]{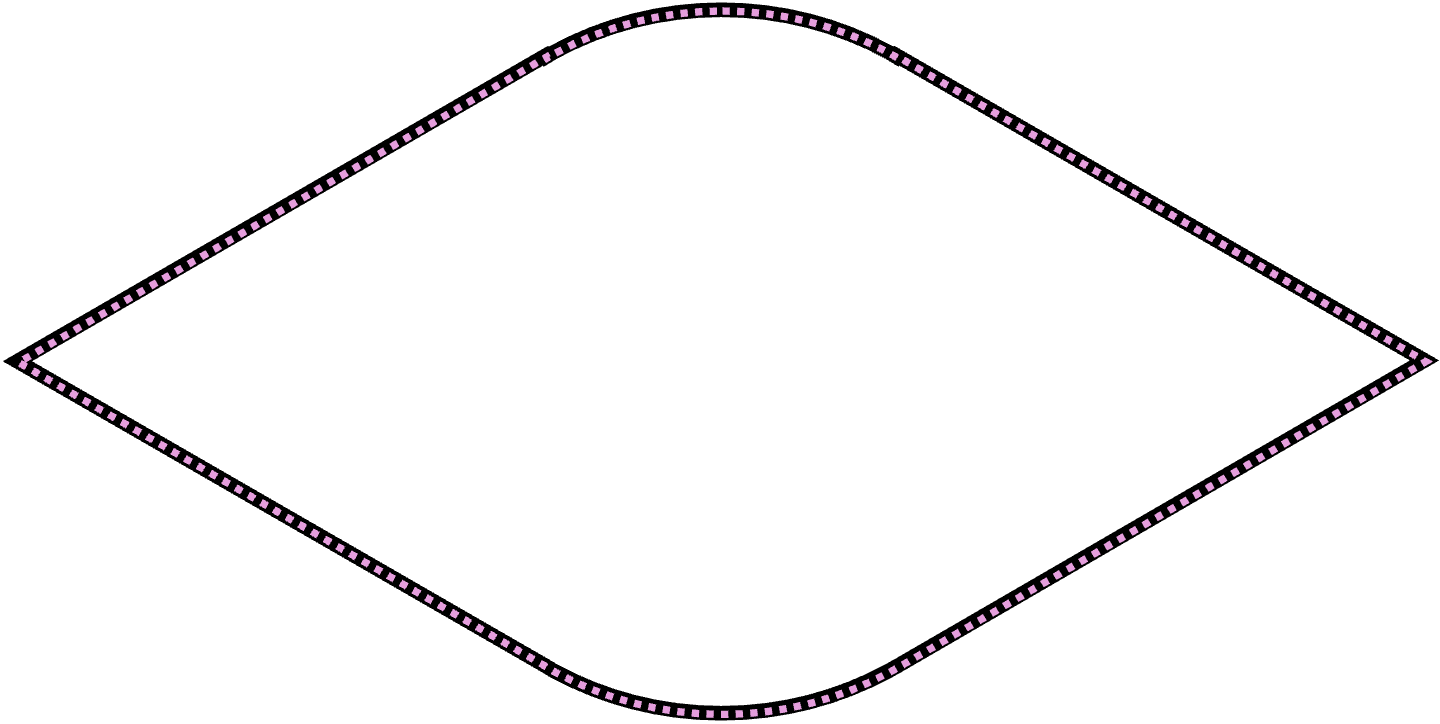}} &
  \parbox[c][1.5cm][c]{2.45cm}{\centering\includegraphics[width=\linewidth]{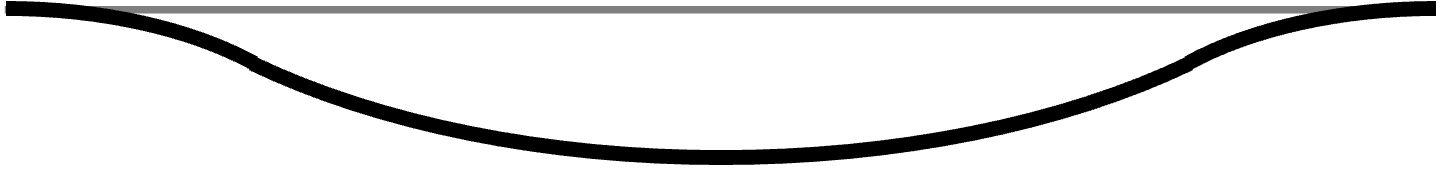}} &
  \parbox[c][1.5cm][c]{2.45cm}{\centering\includegraphics[width=\linewidth]{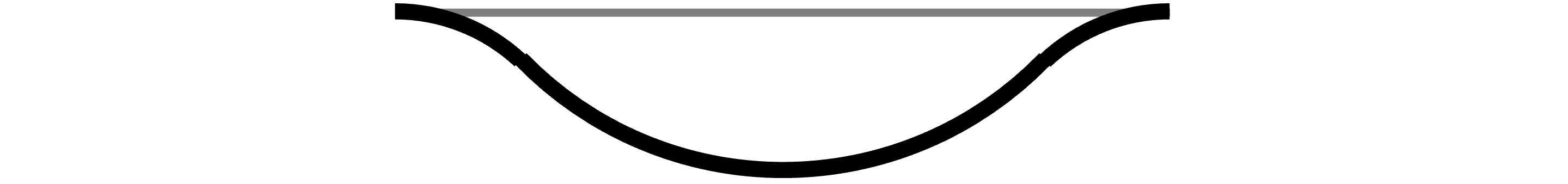}} \\
$\mathbbold{5}$ & Triangle-down &
  \parbox[c][1.5cm][c]{2.45cm}{\centering\includegraphics[width=\linewidth]{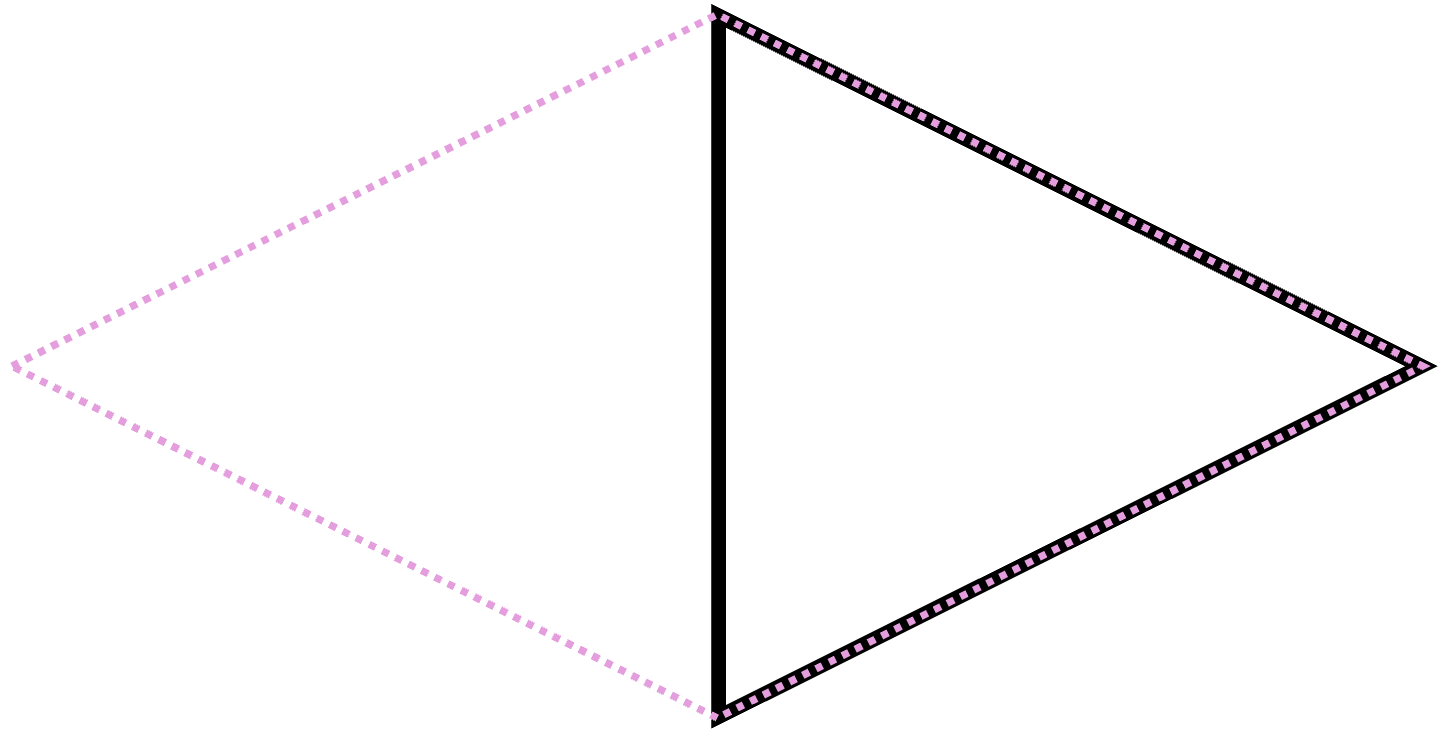}} &
  \parbox[c][1.5cm][c]{2.45cm}{\centering\includegraphics[width=\linewidth]{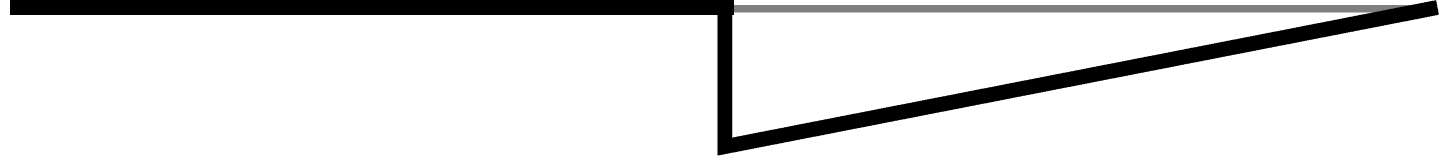}} &
  \parbox[c][1.5cm][c]{2.45cm}{\centering\includegraphics[width=\linewidth]{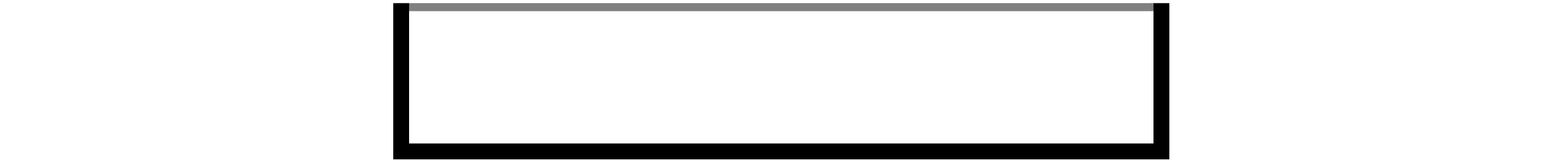}} \\
$\mathbbold{6}$ & Triangle-up &
  \parbox[c][1.5cm][c]{2.45cm}{\centering\includegraphics[width=\linewidth]{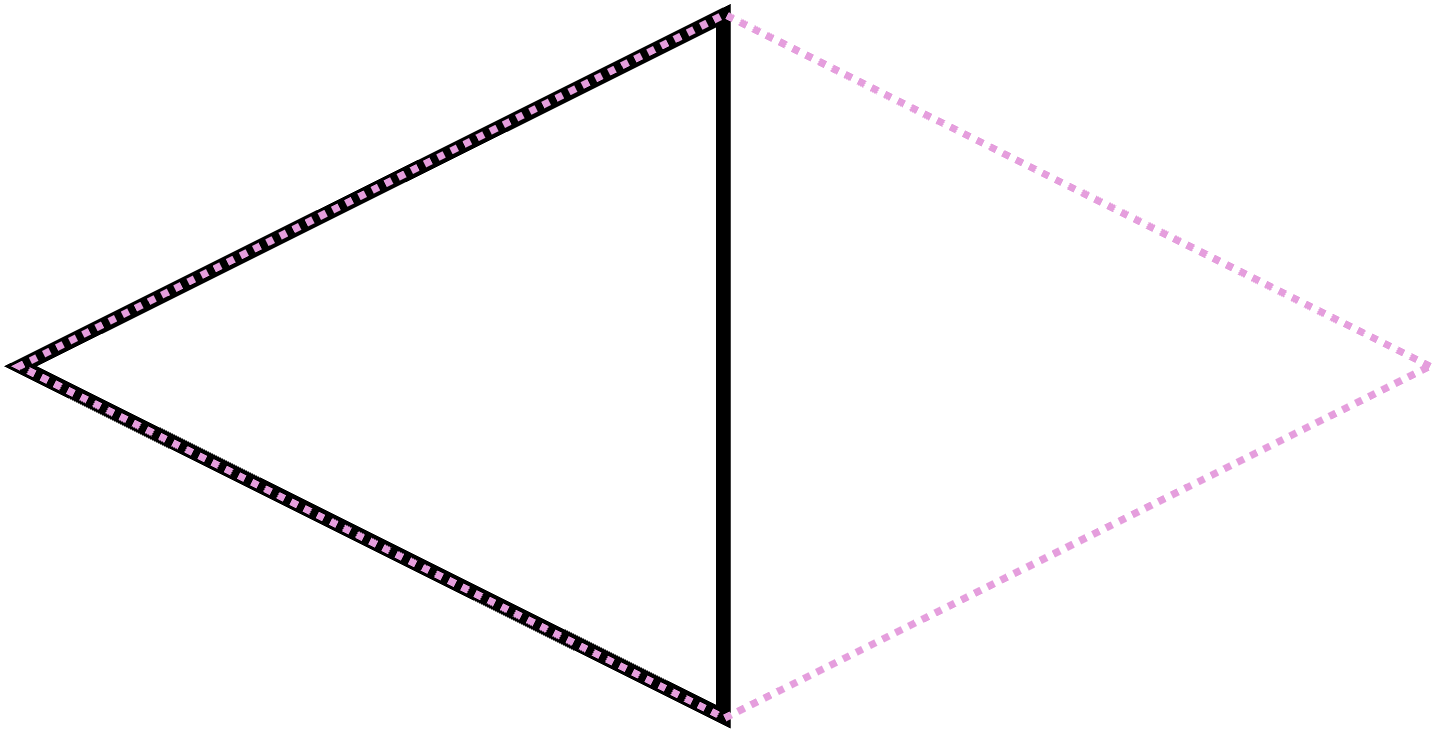}} &
  \parbox[c][1.5cm][c]{2.45cm}{\centering\includegraphics[width=\linewidth]{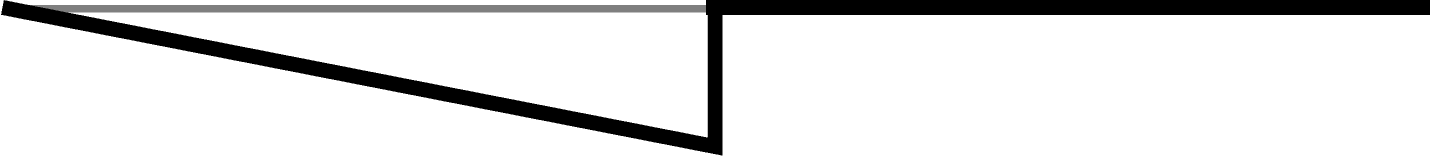}} &
  \parbox[c][1.5cm][c]{2.45cm}{\centering\includegraphics[width=\linewidth]{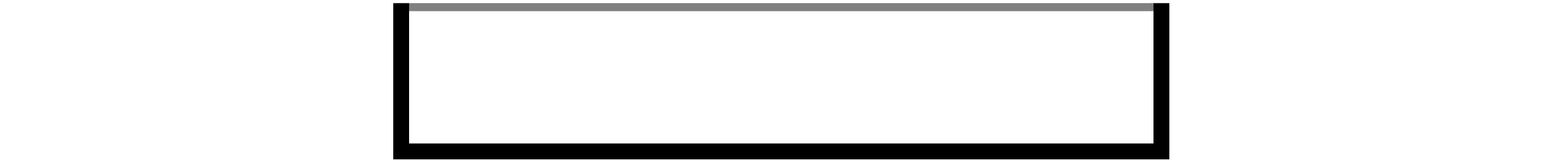}} \\
\hline\hline
\end{tabularx}
\end{table}

\begin{figure}[t]
    \centering
    \includegraphics[width=1\linewidth]{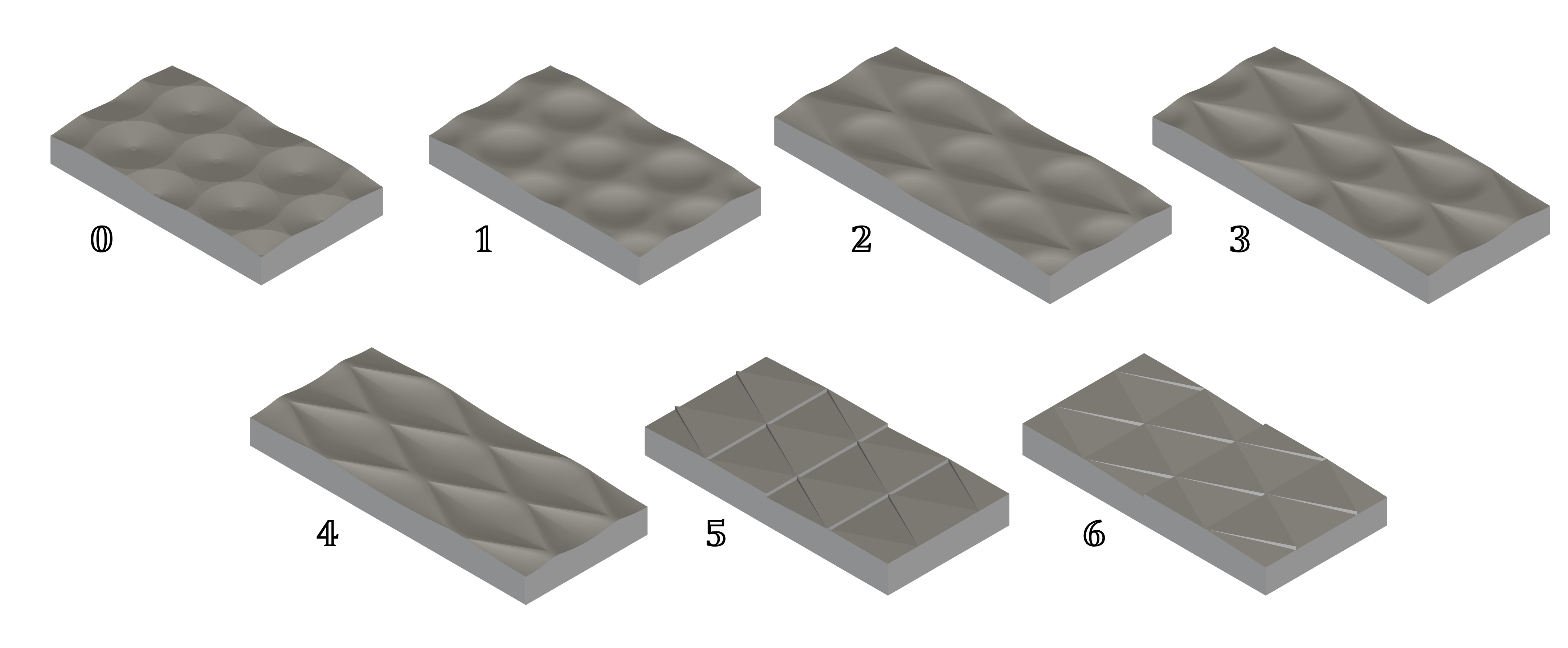}
    \caption{Perspective three-dimensional STL views of the nominal baseline dimple topographies.}
\label{fig:baseline_stl}
\end{figure}

To ensure geometric consistency for the proposed design space, all baseline dimple geometries are generated using an automated parametric geometry-generation pipeline developed in Rhinoceros through RhinoScript. The generated CAD surfaces are subsequently tessellated and exported in STL format using a maximum edge length of $D/100$ for meshing, where $D$ denotes the absolute dimple unit size (Figure~\ref{fig:surface_schema}). This resolution provides an accurate representation of the prescribed dimple geometry while maintaining a computationally efficient triangular surface discretization. The resulting STL models constitute watertight manifold surfaces with consistently oriented face normals and are directly employed as the geometric input to the immersed-boundary LES framework. In Figure~\ref{fig:baseline_stl}, perspective three-dimensional views of the nominal baseline topologies are rendered from the STL tessellations. This data structure directly supplies the precise solid-domain description required by the immersed-boundary preprocessor, ensuring an accurate geometric transfer into the CFD flow solver (see Section~\ref{sec2.2}).

Before elaborating on the geometric controls ($d$, $s$ and $g$), we leave notes on two additional reference parameters anchoring the physical dimensions of the dimpled surface: a fundamental aspect ratio $\gamma$ and an absolute dimple unit size $D$ (see Figure~\ref{fig:surface_schema}). $\gamma$ defines the essential streamwise-to-spanwise length proportions of a baseline dimple unit, ensuring that when the in-plane scale parameter is un-stretched ($s = 1.0$), the geometries exactly recover the nominal configurations assessed in prior literature~\citep{Ng2020, tay2016triangular}; here, $\gamma = 1.0$ for $\tau = \mathbbold{0},~\mathbbold{1}$ (circular planform), and $\gamma = 2.0$ otherwise. $D$ is utilized to scale the physical dimple geometry with the turbulent channel flow field under a constant flow rate condition. Following~\citet{Ng2020}, $D$ is set to five times the channel half-height $\delta$ ($D=5\delta$), while $\delta$ corresponds to $180$ viscous wall units (i.e., $\delta = 180\,\nu/u_{\tau_0}$, where $u_{\tau_0} \equiv (\tau_{w_0} / \rho)^{0.5}$ represents the friction velocity, and $\nu$, $\rho$, and $\tau_{w_0}$ denote the kinematic viscosity, fluid density, and mean wall shear stress of the flat channel, respectively), or equivalently $Re_{\tau} \equiv u_{\tau_0} \delta / \nu = 180$. This condition keeps our designs directly comparable with that baseline and is high enough for the layered structure of wall turbulence to be established~\citep{Ng2020}, which supports extending our findings toward the higher Reynolds numbers in future work.

The three continuous geometric parameters regulate the physical dimensions and spatial layout of the dimples. First, dimple depth $d^+$ is the maximum wall-normal extent of the concavity, where the superscript $+$ indicates the wall-unit normalization ($d^+ \equiv d / (\nu / u_{\tau_0})$). Next, the planform scale $s$ acts as an in-plane multiplier controlling both the dimple size and center-to-center spacing; a setting of $s = 1.0$ defines the full coverage limit where neighboring dimple rims achieve tangential packing (as in Figure~\ref{fig:baseline_stl}). Lastly, the stretch parameter $g$ governs the streamwise elongation of the planform and layout pitch relative to the spanwise direction. It is worthwhile to note that, when applied to the spherical profile ($\tau = \mathbbold{1}$), $g = 1.5$ recovers the elliptical dimple baseline evaluated by~\citet{Ng2020}.

The dimple design optimization problem in the present study, to find the optimal drag-reduction rate denoted $\mathrm{DR}$, is formally posed as:
\begin{align} 
\mathbf{x}_{\mathrm{opt}} &= \underset{\mathbf{x}\in\mathcal{X}}{\arg\max}\; \mathrm{DR}(\mathbf{x}) \label{eq:optimization_problem} 
\end{align} 
where $\mathbf{x} \in \mathcal{X}$ is the mixed-variable design vector; here $\mathbf{x}=(\tau,d^{+},s,g)$ and $ \mathcal{X}$ represents the dimple design space $\mathcal{T}\times\mathcal{D}\times\mathcal{S}\times\mathcal{G}$.
Table~\ref{tab:ordinal-levels} outlines the physical ranges under consideration and the corresponding discrete design levels that define the comprehensive design space.
This framework represents a significant expansion over the preceding DNS investigation by~\citet{Ng2020}, which was restricted to isolated topographies defined at $(d^+,s,g) = (45,1.0,1.0)$. While the present setup encompasses these literature baselines to facilitate direct validation, it simultaneously maps broader geometric variations within a physically justified domain. The upper boundary of the depth range is capped at $d^+ = 50$ to preserve the shallow nature of the dimples; at this maximum setting, the dimple depth corresponds to just over a quarter of the characteristic flow length scale, $\delta$, ensuring the concavities remain a near-wall texture rather than a macro-structural flow obstruction. The resulting design space contains $|\mathcal{X}|=7\times9\times5\times5=1575$ candidate configurations. Since the flat channel reference $G_m^{\mathrm{flat}}$ is fixed, maximizing the drag-reduction rate $\mathrm{DR}$ (see Section~\ref{sec2.2.2} for definition) is equivalent to minimizing $\left|G_m(\mathbf{x})\right|$.

We note here that although $d^{+}$, $s$, and $g$ admit continuous physical definitions, they are represented here by finite ordered level sets, while the dimple topology $\tau$ remains an unordered categorical variable. This mixed-variable discretization is motivated by both computational and optimization considerations. In probabilistic global optimization, the exploration--exploitation trade-off can lead to excessive sampling within narrow neighborhoods, particularly when the response is high-frequency or affected by numerical noise \cite{sheikh2022optimization}. For an objective supplied by computationally expensive wall-resolved LES, such locally clustered evaluations provide limited additional information while consuming a substantial portion of the available budget. Restricting the geometric parameters to discrete levels mitigates this behavior by excluding arbitrarily close candidates and encouraging the optimizer to evaluate distinct regions of the admissible design space \cite{Lee2024}.

\begin{table}[t]
\caption{Summary of physical ranges and design levels for the dimpled surface design variables.}
\label{tab:ordinal-levels}
\centering
\begin{tabularx}{\linewidth}{ccXX} 
\hline\hline
\textbf{Variable} & \textbf{Physical Range} & \textbf{Design Set} \\
\hline
\parbox[c][.75cm][c]{2.45cm}{\centering Depth $d^+$} & 
\parbox[c][.75cm][c]{2.45cm}{\centering $10 \leq d^+ \leq 50$} & 
$\mathcal{D}=\left\{10, 15, 20, 25, 30, 35, 40, 45, 50\right\}$ \\
\parbox[c][.75cm][c]{2.45cm}{\centering Scale $s$} &
\parbox[c][.75cm][c]{2.45cm}{\centering $0 < s \leq 1$} & 
$\mathcal{S}=\left\{0.2, 0.4, 0.6, 0.8, 1.0\right\}$ \\
\parbox[c][.75cm][c]{2.45cm}{\centering Stretch $g$} & 
\parbox[c][.75cm][c]{2.45cm}{\centering $0.5 \leq g \leq 1.5$} &
$\mathcal{G}=\left\{0.5, 0.75, 1.0, 1.25, 1.5\right\}$ \\
\hline\hline
\end{tabularx}
\end{table}

\subsection{Flow Solver}\label{sec2.2}

\subsubsection{Governing Equations and Boundary Conditions}
The aerodynamic performance of each dimpled surface design is evaluated through the in-house LES solver, which is built upon the immersed-boundary (IB)-LES formulation of~\citet{Lee2019}. Given a reference velocity scale $U_{\mathrm{ref}}$ and a reference length scale $L_{\mathrm{ref}}$, the grid-filtered incompressible Navier--Stokes equations are expressed as follows:
\begin{equation}
\frac{\partial \bar{u}_i}{\partial x_i} - m = 0,
\label{eq:continuity}
\end{equation}
\begin{equation}
\frac{\partial \bar{u}_i}{\partial t}
  + \frac{\partial (\bar{u}_i \bar{u}_j)}{\partial x_j}
  = -\frac{\partial \bar{p}}{\partial x_i}
  + \frac{1}{Re}\frac{\partial^2 \bar{u}_i}{\partial x_j \partial x_j}
  - \frac{\partial \tau_{ij}}{\partial x_j}
  + f_i,
\label{eq:momentum}
\end{equation}
where $\bar{u}_i$ and $\bar{p}$ denote the grid-filtered velocity and pressure, respectively; $\tau_{ij}\equiv\overline{u_i u_j}-\bar{u}_i\bar{u}_j$ represents the subgrid-scale (SGS) stress tensor closed by an eddy viscosity model; $f_i$ is the IB forcing that enforces the no-slip condition on embedded surfaces; and $m$ is a mass source/sink term associated with the IB correction. The Reynolds number is defined as $Re\equiv U_{\mathrm{ref}}L_{\mathrm{ref}}/\nu$. Time advancement follows a semi-implicit fractional-step scheme, using a third-order Runge--Kutta method for the convective terms and the Crank--Nicolson scheme for the diffusive terms, with the time step dynamically adjusted to maintain a Courant number of unity. The spatial discretization employs a second-order central-difference method.

When all energetically relevant scales are resolved on a sufficiently fine mesh, Eqs.~\eqref{eq:continuity} and~\eqref{eq:momentum} reduce to direct numerical simulation (DNS). In LES mode, the unresolved scales are represented through $\tau_{ij}$, commonly closed with eddy viscosity models~\citep{Smagorinsky1963,Vreman2004,Park2006}. We employ the SGS closure of~\citet{Vreman2004} with a model coefficient of $0.01$ (selected \textit{a posteriori} through the flat channel and baseline validation of Section~\ref{sec2.2.4}), because its eddy viscosity vanishes for flow types with zero theoretical subgrid-scale dissipation (e.g., laminar or near-wall regions), which offers good physical reliability in the present wall-resolved setting. While resolving the near-wall topography and the momentum exchange directly, we take advantage of LES by coarsening the mesh toward the channel centerline, away from the walls, which yields significant computational savings relative to a DNS-level mesh.

Dimpled surface topographies are represented on a staggered Cartesian grid using
the finite-volume IB method of~\citet{Kim2001}, as implemented in~\citet{Lee2019}.~\citet{Lee2024} applied the same IB-LES foundation to riblet surfaces; here we retain that foundation but target dimple geometries rather than riblets. The STL topography information (see Section~\ref{sec2.1}) is transferred to the IB preprocessor, which assigns no-slip forcing on the embedded boundary nodes without analytic surface differentiation.

\begin{figure}[t]
    \centering
    \includegraphics[width=0.6\linewidth]{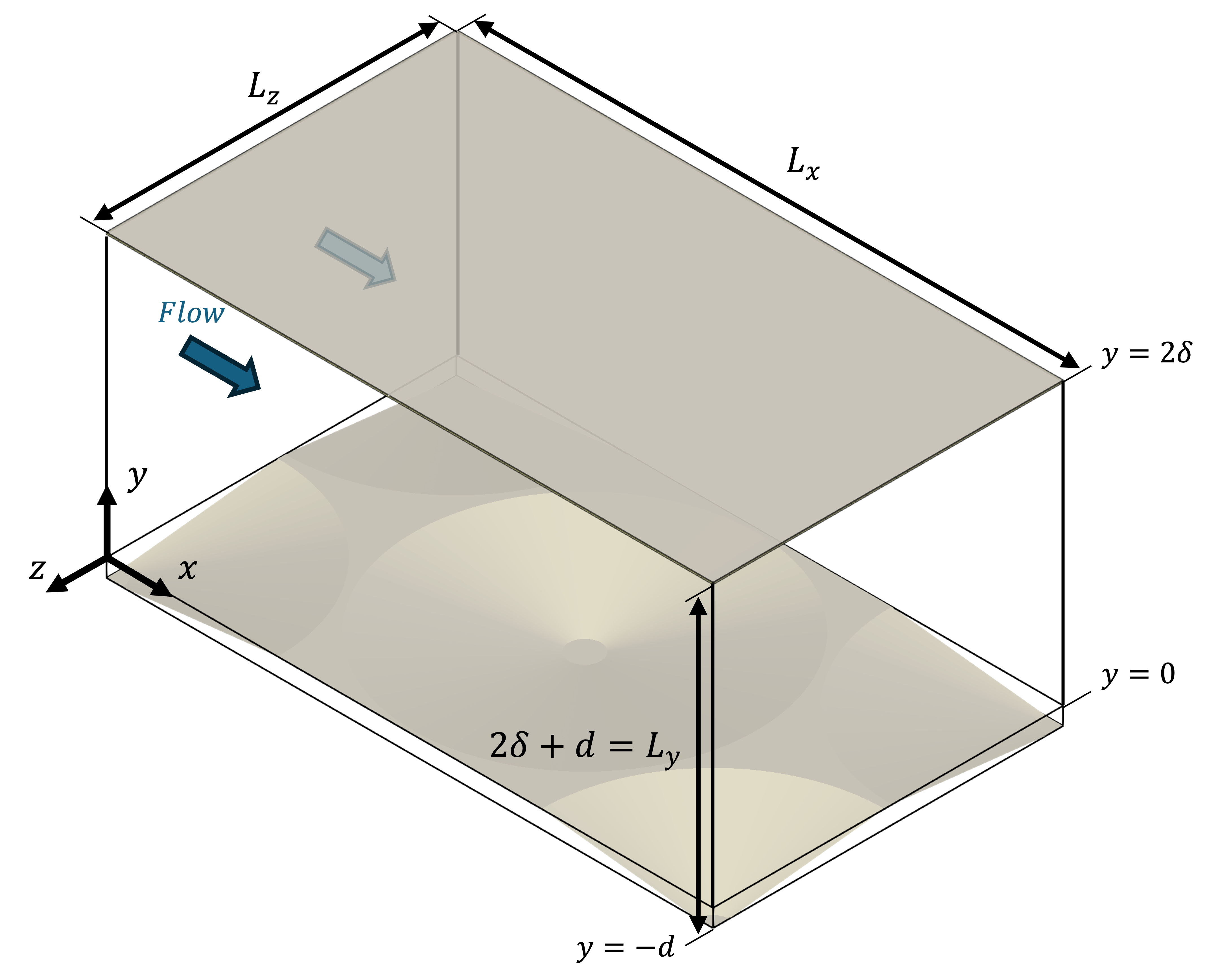}
    \caption{Schematic of the turbulent channel flow configuration.}
    \label{fig:channel_schematic}
\end{figure}

Figure~\ref{fig:channel_schematic} introduces the computational domain setup considered in the present study. For nondimensionalization, $L_{\rm ref}$ is taken to be the flat channel half-height $\delta$. The velocity scale is built on the bulk velocity as the volume-averaged streamwise velocity~\citep{Ng2020},
\begin{equation}
\langle U\rangle \equiv \frac{1}{V}\iiint_\mathcal{F} \bar{u}_1\,\mathrm{d}V,
\label{eq:bulkvel}
\end{equation}
where $V\equiv\iiint_{\mathcal{F}}\mathrm{d}V$ is the volume of the fluid domain $\mathcal{F}$; the reference velocity scale $U_{\mathrm{ref}}$ is then taken to be the flat channel value of $\langle U\rangle$ (denoted $U_b$). These reference scales ensure that all designs, despite their differing dimple geometries, are simulated at a consistent $Re$. The domain is periodic in the streamwise ($x$) and spanwise ($z$) directions, with the domain lengths $L_x$ and $L_z$ chosen to contain exactly one full spatial period of the dimple array.

Because streamwise periodicity cannot accommodate a net pressure drop across the domain, we decompose the filtered pressure explicitly into a spatially uniform mean pressure gradient and a residual component that satisfies periodicity, that is,
\begin{equation}
\bar{p}(x,y,z,t) = \tilde{p}(x,y,z,t) + G_m(t)\,x,
\label{eq:p-decomp}
\end{equation}
where $\tilde{p}$ denotes the residual pressure and $G_m(t)$ the domain-averaged mean pressure gradient, generally negative so as to drive the flow in the positive $x$ direction. The $G_m$ term thus acts as a spatially uniform streamwise momentum source. Under the constant-flow-rate (CFR) condition adopted here, $G_m$ is not prescribed \textit{a priori} but is updated from the wall and IB momentum exchange at every time step: a proportional--integral feedback controller adjusts $G_m$ from the difference between the measured flow rate and the CFR target, so that $G_m$ serves as the controlled driving force that maintains the target flow rate over time.

A no-slip condition is enforced at both channel walls through IB forcing. On the lower wall, we impose an immersed slab carrying the dimpled surface, with its flat rim plane at $y=0$ and its concavities extending into $y<0$; on the upper wall, we represent the plane boundary by a thin, flat immersed slab whose fluid--solid interface lies at $y=2\delta$. The wall-normal domain height thus becomes $L_y = 2\delta + d$ (Figure~\ref{fig:channel_schematic}).

It should be noted that we deliberately retain the flat channel half-height $\delta$ as the fixed reference length for nondimensionalization. The dimples nevertheless extend fluid pockets into $y<0$, enlarging the ``effective'' channel height, i.e., the wall-normal separation between the surface-averaged location of the dimpled lower boundary and the opposite plane wall. To take this into account, following~\citet{Ng2020}, we enforce the CFR condition by holding $\iiint_{\mathcal{F}}\bar{u}_1\,\mathrm{d}V=\langle U\rangle\,V$ fixed at its flat channel value; this entails a compensating reduction of $\langle U\rangle$ slightly below the reference scale $U_b$, which is necessary to preserve effectively the same bulk Reynolds number across the different dimpled configurations.

\subsubsection{Drag Reduction Objective}\label{sec2.2.2}
From the simulations conducted under the CFR condition established above, we evaluate the drag-reduction objective in terms of the mean pressure gradient $G_m$. For statistically stationary, fully developed channel flow, the time-averaged $G_m$ balances the total drag per unit fluid volume. Accordingly, the design optimization run aims to maximize $G_m$ (equivalently, to minimize $|G_m|$), and we quantify drag-reducing performance relative to the flat channel reference evaluated under identical CFR conditions through the drag-reduction rate
\begin{equation}
\mathrm{DR} = \frac{|G_m^{\mathrm{flat}}| - |G_m|}{|G_m^{\mathrm{flat}}|}\times 100\%
\label{eq:DR}
\end{equation}
where $\mathrm{DR}>0$ denotes a favorable net drag reduction. 

It is worth noting that this objective formulation follows that of ~\citet{Ng2020} and obviates any post-processing of the flow field, since $G_m$ is a scalar quantity computed and applied directly as a momentum source during the main solver iterations. However, the total drag (denoted $C_d$) can still be computed directly by volume-integrating the IB forcing over the immersed solid bodies with appropriate corrections~\citep[see][]{Lee2024}; we adopt this direct evaluation secondarily as an independent cross-check against $G_m$ that confirms the force balance and the adequacy of the time-averaging window.

\subsubsection{Computational Mesh}\label{sec2.2.3}
In this section, all grid spacings are reported in viscous wall units (``$+$'' notations; see Section~\ref{sec2.1}) built on the flat channel friction velocity $u_{\tau_0}$ and half-height $\delta$; we hold this ``$+$'' scaling fixed at the flat channel reference across every design, so that although the local wall shear varies from case to case, the resolution is assessed against a consistent normalization. 

In the wall-parallel directions, we maintain uniform streamwise and spanwise spacings of $\Delta x^+ = 10$ and $\Delta z^+ = 10$, kept fine to capture the dimple topography. Since $L_x$ and $L_z$ span one full period of the dimple array, the wall-parallel grid count scales with the pattern size. In the wall-normal direction, every cell in $y<0$ or adjacent to no-slip surface is held at $\Delta y^+ = 1$, as required for wall-resolved near-wall physics~\citep{Davidson2009}, and is then gradually stretched toward the channel center, where the LES spacing reaches $\Delta y^+ = 12$. In case we perform the DNS re-evaluations, the mesh is refined by roughly a factor of two in each direction ($\Delta x^+ = 5$, $\Delta z^+ = 5$ with a centerline $\Delta y^+ = 5$), reaching a DNS-level resolution~\citep{Kim1987}.

\subsubsection{Simulation Setup and Validation}\label{sec2.2.4}
\begin{figure}[t]
    \centering
    \includegraphics[width=0.72\linewidth]{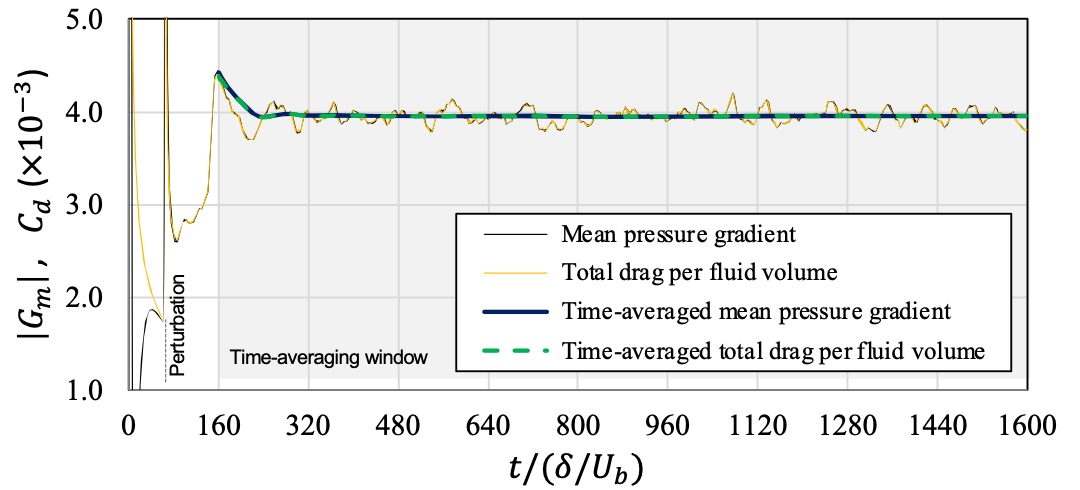}
    \caption{Time histories of the mean pressure gradient $|G_m|$ and the directly
    integrated total drag per fluid volume $C_d$ in the $Re_\tau = 180$ flat channel LES.}
    \label{fig:gm_history}
\end{figure}

Each simulation is initialized from rest. To bring the quiescent fluid up to the desired flow rate, the $G_m$ controller first ramps up a large mean pressure gradient; once the bulk velocity reaches the CFR target, $G_m$ is regulated by the proportional--integral feedback so that the CFR is maintained thereafter. Transition to turbulence is activated at $t = 64$ (n.b. nondimensionalized by $t_{\rm ref} \equiv L_{\rm ref} / U_{\rm ref} = \delta / U_b$) by superposing a finite-amplitude sinusoidal perturbation on the developing fluid, and the flow settles into a statistically stationary turbulent state well before averaging begins. We accumulate flow statistics over the long window from $t = 160$ to $t = 1600$. Figure~\ref{fig:gm_history} shows the time histories of $|G_m|$ and of the directly calculated total drag per fluid volume $C_d$ in the flat channel LES (see below). Both plateau to a common value long before $t = 1600$, confirming that the chosen window yields a converged, statistically stationary force balance. We apply the same averaging window to every dimpled design; each case's $G_m$ history reaches a comparable statistically stationary plateau within it, including the designs that develop near-wall separation.

\begin{figure}[t]
    \centering
    \includegraphics[width=0.8\linewidth]{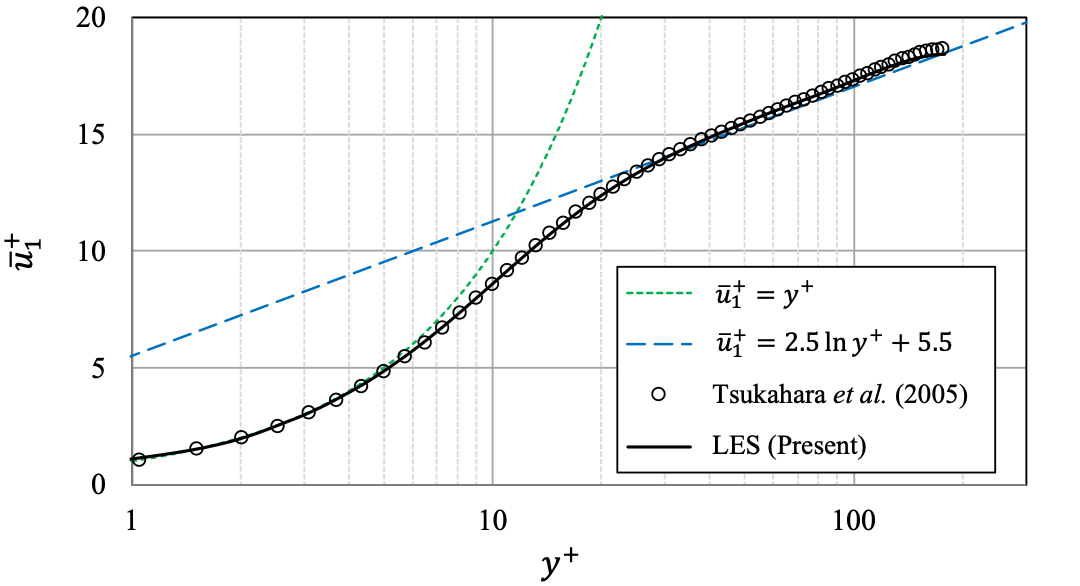}
    \caption{Mean streamwise velocity profile $\bar{u}_1^+(y^+)$ of the present $Re_\tau=180$ flat channel LES (black line), compared with the reference DNS data of~\citet{Tsukahara2005} (circles).}
    \label{fig:uplus}
\end{figure}

To establish the trust of our LES solver, we conduct validation checks against published high-fidelity data. First, we validate the solver on the flat channel at $Re_\tau = 180$; the computational domain is set to $L_x =10\delta$ and $L_z = 5\delta$ (or $L_x ^+ = 1800$ and $L_z^+ = 900$), which is sufficiently larger than the minimal flow unit suggested by~\citet{Jimenez1991} to yield meaningful one-point turbulent statistics. The flat channel attains $|G_m^{\rm flat}| = 3.956 \times 10^{-3}$  (as shown in Figure~\ref{fig:gm_history}) at a bulk Reynolds number $Re = U_b \delta / \nu = 2845$. Using the fact that $Re_\tau / Re = u_\tau / U_b = (\tau _{w_0} / (\rho U_b^2))^{1/2} = |G_m^{\rm flat}|^{1/2}$, in which the last equality assumes the force balance, the ``computed'' $Re_\tau$ (from the evaluated $|G_m^{\rm flat}|$) is $1.79 \times 10^2$, which is within $0.6\%$ of the ``nominal'' $Re_{\tau}$ of $180$. The bulk Reynolds number $Re = 2845$ is consistent with the canonical DNS data at $Re_\tau = 180$ including \citet{Kim1987} ($Re = 2810$) and \citet{Tsukahara2005} ($Re = 2865$). Figure~\ref{fig:uplus} compares the computed mean velocity profile with the reference DNS of~\citet{Tsukahara2005}, which confirms that the present LES correctly reproduces the viscous sublayer ($\bar{u}_1^+ = y^+$) and recovers the logarithmic profile ($\bar{u}_1^+ = 2.5 \ln y^+ + 5.5$) and centerline value, closely tracking the DNS.

Second, we further validate the LES solver on non-trivial topographies by reproducing the five dimpled surface baselines of~\citet{Ng2020}, each of which coincides with a configuration in our design space. As summarized in Table~\ref{tab:baseline}, we compare, for every baseline, the drag-reduction rate $\mathrm{DR}$ from the present LES against its DNS counterpart. The agreement is consistently close: the LES $\mathrm{DR}$ differs from the DNS by less than 2 percentage points across all five topographies, comfortably below the 3 percentage point simulation-versus-experiment $\mathrm{DR}$ differences regarded as reasonable agreement in the riblet drag reduction research~\citep{Lee2024}. The LES also reproduces the same drag-reduction performance ranking, identifying the diamond design $(\tau,d^+,s,g)=(\mathbbold{4},45,1.0,1.0)$ as the best and the spherical one $(\tau,d^+,s,g)=(\mathbbold{1},45,1.0,1.0)$ as the worst. The fact that the LES recovers the drag response of these complex dimpled surface topographies, beyond the flat channel statistics, supports the solver's physical reliability across the non-trivial topographies populating the dimple design space.

\begin{table}[tb]
\caption{Validation of the present LES against the DNS of~\citet{Ng2020}.}
\label{tab:baseline}
\centering
\begin{tabularx}{\linewidth}{>{\centering\arraybackslash}X c c c c c c c}
\hline\hline
\textbf{Baseline} & \multirow{2}{1em}{$\bm{\tau}$} & \multirow{2}{1em}{$\bm{d^+}$} & \multirow{2}{1em}{$\bm{s}$} & \multirow{2}{1em}{$\bm{g}$} & \multicolumn{2}{c}{\textbf{DR (\%)}} & \multirow{2}{4em}{\textbf{Rank}}\\
\textbf{Designs}& & & & & \textbf{LES (Present)} & \textbf{\citet{Ng2020}} & \\
\hline
Spherical        & \parbox[c][.75cm][c]{1em}{$\mathbbold{1}$} & 45 & 1.0 & 1.0 & $-7.51$ & $-6.4$ & 5 (worst) \\
Elliptical       & \parbox[c][.75cm][c]{1em}{$\mathbbold{1}$} & 45 & 1.0 & 1.5 & $+3.74$ & $+4.9$ & 2 \\
Teardrop-down    & \parbox[c][.75cm][c]{1em}{$\mathbbold{2}$} & 45 & 1.0 & 1.0 & $+0.94$ & $-0.1$ & 4 \\
Teardrop-up      & \parbox[c][.75cm][c]{1em}{$\mathbbold{3}$} & 45 & 1.0 & 1.0 & $+2.76$ & $+3.1$ & 3 \\
Diamond          & \parbox[c][.75cm][c]{1em}{$\mathbbold{4}$} & 45 & 1.0 & 1.0 & $+5.61$ & $+7.4$ & 1 (best) \\
\hline\hline
\end{tabularx}
\end{table}

\subsection{Bayesian Optimization}
\label{sec2.3}

Bayesian optimization (BO) is particularly well suited to engineering design problems in which the objective is available only through computationally expensive black-box evaluations. Rather than relying on large populations or exhaustive exploration, BO constructs a probabilistic surrogate of the objective and uses an acquisition function to determine which design should be evaluated next. This enables the search to balance the exploitation of designs predicted to perform well against the exploration of poorly resolved regions of the design space \citep{brochu2010tutorial}. In the present problem, every objective evaluation requires a statistically stationary, wall-resolved large eddy simulation of turbulent channel flow. Moreover, the response of a dimpled surface can vary nonlinearly with topology, depth, coverage, and elongation, and potentially contains numerical and sampling uncertainty associated with finite-time turbulence statistics. An exhaustive evaluation of all $1575$ candidate configurations would therefore be prohibitively expensive, while a purely local or greedy search could overlook favorable combinations arising from interactions among the design variables. BO provides a sample-efficient framework for identifying drag-reducing configurations while limiting the number of high-fidelity flow simulations. 

The optimization is performed using the Mixed-variable Multi-Objective Bayesian Optimization (MixMOBO) framework \citep{Sheikh2022a}. The framework is employed here in a single-objective setting to maximize the drag-reduction rate. Algorithm~\ref{alg:mixmobo} summarizes the MixMOBO procedure.  The design vector $\mathbf{x}=(\tau,d^{+},s,g)$ is inherently mixed-variable: the dimple topology $\tau$ is an unordered categorical variable, whereas depth $d^{+}$, scale $s$, and stretch $g$ are ordered geometric variables defined on the discrete level sets listed in Table~\ref{tab:ordinal-levels}. MixMOBO accommodates these variable types within a common Gaussian-process-based optimization framework, avoiding the artificial assumption that successive topology labels possess a meaningful numerical distance. Specifically, the Gaussian process covariance employs categorical mismatch for $\tau$ and numerical distance for the ordered geometric variables. A Mat\'ern-$5/2$ kernel is used, with its hyperparameters determined from the available simulation data through leave-one-out cross-validation. The discretization of the geometric variables also prevents the optimizer from repeatedly evaluating arbitrarily close, and consequently nearly redundant, configurations. This is advantageous for expensive and potentially noisy response surfaces, for which conventional BO may concentrate an excessive number of evaluations within small neighborhoods \citep{Lee2024,sheikh2022optimization}. 

To improve robustness against the behavior of any single acquisition function, the in-built Hedge strategy is used to manage a portfolio consisting of expected improvement, probability of improvement, Gaussian process upper confidence bound, and Thompson sampling. At each optimization epoch, these acquisition functions propose competing candidates representing different exploration--exploitation preferences. Each acquisition function is maximized over the unevaluated mixed-variable design space using a genetic search, and the hedging mechanism selects one of the resulting nominees according to probabilities determined by the accumulated acquisition gains \citep{Sheikh2022a}. The selected design is then evaluated using LES, its drag-reduction rate is appended to the training data, and the Gaussian process surrogate and Hedge gains are updated before the next optimization epoch. 

The optimization is initialized with $50$ designs, including the five reference configurations of \citet{Ng2020}, with the remaining $45$ designs randomly sampled from the admissible design space $\mathcal{X}$. This initial set establishes broad coverage of the mixed design space while anchoring the search to previously studied dimple topographies. The initial evaluations are followed by $100$ MixMOBO-guided LES evaluations, giving a total budget of $150$ designs. Thus, approximately one third of the simulation budget is allocated to initial exploration, while the remaining two thirds are used for adaptive sampling of promising and uncertain regions. This allocation supplies the surrogate with a sufficiently diverse initial data set while preserving the majority of the computational budget for BO-guided refinement. The final optimized design is taken as the evaluated configuration with the largest drag-reduction rate over the complete set of $150$ LES evaluations.

\begin{algorithm}[!tb] 
\caption{MixMOBO for Optimizing Dimple Design for Drag Reduction} 
\label{alg:mixmobo} 

\KwIn{Design space $\mathcal{X}$, initial sample size $N_0$, total budget $N$, acquisition portfolio $\mathcal{A}$} 


Sample the initial designs $\mathbf{X}_0=\{\mathbf{x}_i\}_{i=1}^{N_0}\subset\mathcal{X}$ and evaluate $y_i=\mathrm{DR}(\mathbf{x}_i)$ using LES.\;

Initialize $\mathcal{D}_0=\{(\mathbf{x}_i,y_i)\}_{i=1}^{N_0}$ and Hedge gains $g_0^k=0$ for all $\alpha_k\in\mathcal{A}$.\;

\For{$t=N_0,\ldots,N-1$}{ 

Fit the mixed-variable Gaussian process $f(\mathbf{x})\mid\mathcal{D}_t \sim \mathcal{GP}\!\left( \mu_t(\mathbf{x}),\sigma_t^2(\mathbf{x}) \right)$ using a Mat\'ern-$5/2$ kernel with hyperparameters determined by leave-one-out cross-validation.\;

For each $\alpha_k\in\mathcal{A}$, obtain the acquisition nominee $\displaystyle \mathbf{x}_{t+1}^{k} = \arg\max_{\mathbf{x}\in \mathcal{X}\setminus\mathbf{X}_t} \alpha_k\!\left( \mathbf{x};\mu_t,\sigma_t \right)$ using mixed-variable genetic search.\;

Compute the Hedge selection probabilities $\displaystyle p_t^k = \exp(\eta_t g_t^k) \big / \sum_{j=1}^{|\mathcal{A}|}\exp(\eta_t g_t^j)$ and select $\mathbf{x}_{t+1}\leftarrow\mathbf{x}_{t+1}^{k}$ with probability $p_t^k$.\; 

Evaluate $y_{t+1}=\mathrm{DR}(\mathbf{x}_{t+1})$ using LES and update $\displaystyle \mathcal{D}_{t+1} = \mathcal{D}_t \cup \{(\mathbf{x}_{t+1},y_{t+1})\}$.\;

Update Hedge gains $\{g_{t+1}^k\}_{k=1}^{|\mathcal{A}|}$ from the posterior performance of the acquisition nominees.\; } 

\Return{$\displaystyle \mathbf{x}_{\mathrm{opt}} = \argmax_{(\mathbf{x}_i,y_i)\in\mathcal{D}_N}y_i$}\; 

\end{algorithm}

\section{Results}
\label{sec3}

\subsection{Optimization History and Solutions}
\label{sec3.1}
The results from our optimization run are summarized in Table~\ref{tab:campaign}. The best-performing design is a relatively deep, fully packed, streamwise-elongated diamond topography $(\tau, d^+, s, g) = (\mathbbold{4},50,1.0,1.5)$, with $G_m=-3.433\times10^{-3}$, i.e., $\mathrm{DR}=13.2\%$. Among the assessed 150 dimple design candidates, $95$ ($63\%$) reduce drag relative to the flat wall while $55$ ($37\%$) increase it, confirming that shallow concavities are not unconditionally beneficial in drag reduction. The range of $\mathrm{DR}$ spans $13.2\%$ (best) to $-84.9\%$ (worst). Although the optimization run covers only $9.5\%$ of the design space ($150$ out of all $1575$ design candidates; see Table~\ref{tab:ordinal-levels}), it nonetheless identifies the best-performing LES-verified design at the very early optimization stage (12th BO epoch) that outperforms the baselines of~\citet{Ng2020}.

\begin{table}[t]
\caption{BO run summary for drag-reducing dimpled surface design exploration.}
\label{tab:campaign}
\centering
\begin{tabularx}{\linewidth}{>{\raggedright\arraybackslash}X l}
\hline\hline
\textbf{Quantity} & \textbf{Value} \\
\hline
\# of samples & 150 (50 initial random + 100 BO-guided) \\
Best sample & 62nd (12th BO epoch) \\
Best design & $(\tau,d^+,s,g)=(\mathbbold{4},50,1.0,1.5)$\\
Best $\mathrm{DR}$ & 13.2\% \\
Best $|G_m|$ & $3.433\times10^{-3}$ \\
Drag-reducing dimple designs ($\mathrm{DR}>0$) & 95 / 150 (63\%) \\
Drag-increasing dimple designs ($\mathrm{DR}<0$) & 55 / 150 (37\%) \\
$\mathrm{DR}$ range & $13.2\%$ to $-84.9\%$ \\
\hline\hline
\end{tabularx}
\end{table}

\begin{figure}[t]
    \centering
    \includegraphics[width=0.82\linewidth]{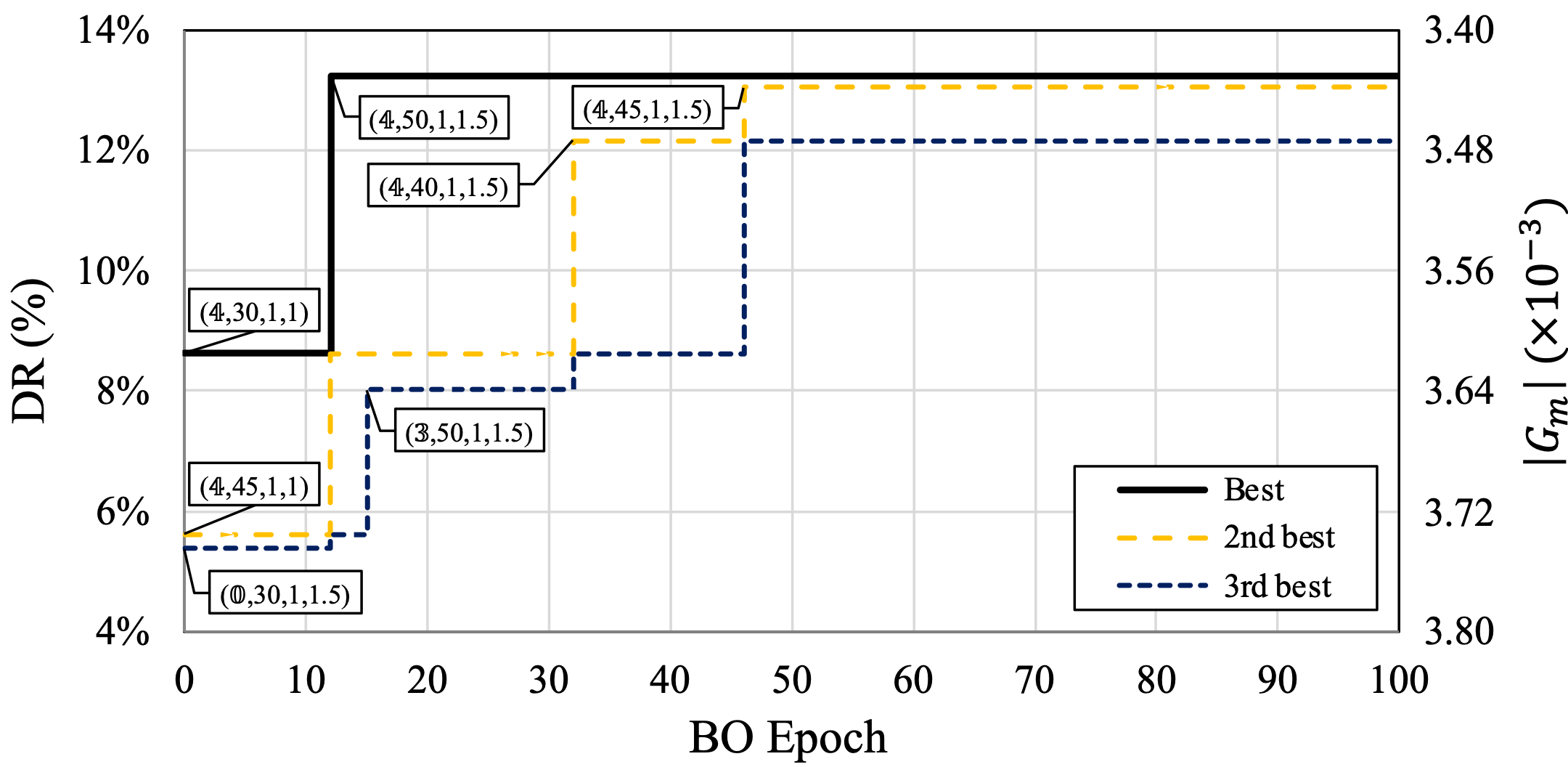}
    \caption{Bayesian optimization (BO) history of the top three drag-reducing dimple designs across BO epochs. The new dimple design entries are annotated by their design variable tuples $(\tau,d^+,s,g)$.}
\label{fig:history}
\end{figure}

The fact that the optimum is discovered early in the BO phase (epoch~12) and is not displaced by the subsequent 88 LES evaluations indicates convergence on a best dimple design. Tracking the best-design incumbent (top curve) in Figure~\ref{fig:history} makes this concrete: it remains within the diamond type ($\tau=\mathbbold{4}$) throughout the BO phase and advances from a moderately deep, unstretched diamond $(\tau,d^+,s,g)=(\mathbbold{4},30,1.0,1.0)$ to the deeper, streamwise-elongated one $(\mathbbold{4},50,1.0,1.5)$, after which it no longer moves. The runner-up traces provide a complementary convergence signal. Their top-3 updates occur at epochs~15, 32, and 46; after epoch~46, no further improved design enters the running top three incumbents through epoch~100. This further supports BO convergence over the final 54 epochs.

Table~\ref{tab:phase} contrasts the BO-guided optimization phase with the initial random sample collection phase, showing that the Bayesian search improves the objective $\mathrm{DR}$ relative to the initial exploration while reducing its standard deviation by a factor of $2.5$. Relative to the flat baseline, $79$ of $100$ BO designs ($79\%$) beat the flat wall, versus only $16$ of $50$ initial designs ($32\%$). Taken together, these statistics indicate that the present MixMOBO scheme meaningfully steers design exploration toward drag-reducing regions of the space as desired.

\begin{table}[t]
\caption{Phase comparison of the objective $\mathrm{DR}$.}
\label{tab:phase}
\centering
\begin{tabularx}{\linewidth}{>{\raggedright\arraybackslash}X c c c c}
\hline\hline
\textbf{Phase} & \textbf{\# of samples} & \textbf{Mean (\%)} & \textbf{Std. (\%)} & \textbf{Best (\%)} \\
\hline
Initial random phase &  \parbox[c][.75cm][c]{1em}{50} & $-5.93$ & $15.4$ & $8.62$ \\
BO-guided phase &  \parbox[c][.75cm][c]{1em}{100} & $2.98$ & $6.12$ & $13.2$ \\
\hline\hline
\end{tabularx}
\end{table}

The five best dimple topography designs discovered in the present run are displayed in Table~\ref{tab:best}. Notably, they are topologically homogeneous: every one is a diamond type ($\tau=\mathbbold{4}$) at full coverage ($s=1.0$) and streamwise-elongated ($g=1.25$ or $g=1.5$), with depths in the upper half of the design range ($d^+=40$--$50$). This strongly suggests that drag reduction, at least within our constructed design space, is set by a specific combination of dimple type, coverage, elongation, and depth rather than by a sole design factor. Across this type the drag-reduction rate varies only modestly, from $11.0\%$ to $13.2\%$, and orders cleanly with the two continuous parameters: at fixed elongation it grows with depth (compare ranks~1 through~3), and at fixed depth it grows with streamwise elongation (compare ranks~3 and~4). The observed optimum, $(\tau,d^+,s,g)=(\mathbbold{4},50,1.0,1.5)$ delivering a $13.2\%$ $\mathrm{DR}$, is the deepest, most elongated, and fully packed configuration in the current dimple surface design space.

To verify the LES $\mathrm{DR}$ outcomes, each design is re-evaluated under the same simulation setup but in DNS mode, where the mesh is refined to DNS-level resolution (${\sim}2\times$ finer in each direction; see Section~\ref{sec2.2.3}) and the SGS model is accordingly deactivated. These DNS runs enlarge the resource requirements by roughly a factor of $2^3$, with the computing time growing proportionately, which is why the DNS check is confined to these best designs. Throughout this check, the LES $\mathrm{DR}$ values are found to reproduce the higher-fidelity DNS outputs closely: for every design the DNS $\mathrm{DR}$ lies within $0.3$ percentage points of its LES value, and the ranking is preserved. This close match supports the credibility of the present LES evaluations.

\begin{table}[t]
\caption{Top five designs discovered from the present optimization run. All five designs are re-evaluated using our solver in the higher-resolution DNS mode (see Section~\ref{sec2.2.3}).}
\label{tab:best}
\centering
\small
\begin{tabular}{cccccccc}
\hline\hline
\textbf{Rank} & $\bm{\tau}$ & $\bm{d^+}$ & $\bm{s}$ & $\bm{g}$ & $\bm{|G_m|\,(\times10^{-3})}$ & \textbf{LES} $\bm{\mathrm{DR}}$ \textbf{(\%)} & \textbf{DNS} $\bm{\mathrm{DR}}$ \textbf{(\%)} \\
\hline
1 & \parbox[c][.75cm][c]{1em}{$\mathbbold{4}$} & $50$&$1.0$&$1.5$ & $3.433$ & 13.2\% & 13.5\% \\
2 & \parbox[c][.75cm][c]{1em}{$\mathbbold{4}$} & $45$&$1.0$&$1.5$ & $3.440$ & 13.0\% & 13.1\% \\
3 & \parbox[c][.75cm][c]{1em}{$\mathbbold{4}$} & $40$&$1.0$&$1.5$ & $3.475$ & 12.2\% & 12.4\% \\
4 & \parbox[c][.75cm][c]{1em}{$\mathbbold{4}$} & $40$&$1.0$&$1.25$ & $3.513$ & 11.2\% & 11.2\% \\
5 & \parbox[c][.75cm][c]{1em}{$\mathbbold{4}$} & $45$&$1.0$&$1.25$ & $3.519$ & 11.0\% & 11.2\% \\
\hline\hline
\end{tabular}
\end{table}

\subsection{Sample-Based Geometric Trends}\label{sec3.2}

Having identified the best dimple designs, we turn to what the collected samples reveal about the geometry of drag-reducing dimples. Table~\ref{tab:topodr} groups the evaluated designs by dimple type ($\tau=\mathbbold{0}$--$\mathbbold{6}$) in order of mean $\mathrm{DR}$. The diamond ($\tau=\mathbbold{4}$) is the clearly favorable type, with $85\%$ of its evaluated designs beating the flat wall. The teardrop types ($\mathbbold{2}$ and $\mathbbold{3}$) are mildly favorable, while the tapered cylinder and spherical ($\mathbbold{0}$ and $\mathbbold{1}$) hover near the flat-wall baseline. The triangles ($\mathbbold{5}$ and $\mathbbold{6}$) are strongly drag-increasing, indicating that a sharp dimple apex is unfavorable for drag reduction, in line with \citet{tay2016triangular}. However, the wide range between the best and worst $\mathrm{DR}$ within each dimple type demonstrates that drag reduction is set not by the mere presence of dimples but by a complex interaction among several geometric factors \citep[e.g.,][]{Ng2020,Nasr2022}.

\begin{table}[t]
\caption{Drag reduction rate by dimple type (\# = number of evaluated designs).}
\label{tab:topodr}
\centering
\begin{tabularx}{\linewidth}{>{\raggedright\arraybackslash}X c c c c c}
\hline\hline
$\bm{\tau}$ (\textbf{Dimple Type}) & \textbf{\#} & \textbf{Mean (\%)} & \textbf{Best (\%)} & \textbf{Worst (\%)} & \textbf{Frac.} $\bm{\mathrm{DR > 0}}$ \\
\hline
$\mathbbold{4}$ (Diamond) & \parbox[c][.75cm][c]{1em}{40} & $5.51$ & $13.2$ & $-14.7$ & 85\% \\
$\mathbbold{2}$ (Teardrop-down) & \parbox[c][.75cm][c]{1em}{19} & $1.89$ & $7.23$ & $-7.51$ & 74\% \\
$\mathbbold{3}$ (Teardrop-up) & \parbox[c][.75cm][c]{1em}{22} & $1.56$ & $8.01$ & $-27.2$ & 77\% \\
$\mathbbold{0}$ (Tapered cylinder) & \parbox[c][.75cm][c]{1em}{21} & $0.65$ & $5.59$ & $-6.17$ & 57\% \\
$\mathbbold{1}$ (Spherical) & \parbox[c][.75cm][c]{1em}{18} & $-1.01$ & $4.27$ & $-12.8$ & 61\% \\
$\mathbbold{5}$ (Triangle-down) & \parbox[c][.75cm][c]{1em}{18} & $-6.07$ & $1.90$ & $-30.0$ & 22\% \\
$\mathbbold{6}$ (Triangle-up) & \parbox[c][.75cm][c]{1em}{12} & $-14.6$ & $1.11$ & $-84.9$ & 25\% \\
\hline\hline
\end{tabularx}
\end{table}

Turning to depth, the aggregate correlation of $\mathrm{DR}$ with $d^+$ across all the sampled designs is weakly negative (i.e., the Pearson correlation coefficient is computed as $r=-0.16$), yet this pooled figure does not reveal type-specific effects that act in opposite directions. Table~\ref{tab:depthslope} shows the fitted local slope of $\mathrm{DR}$ versus depth within each type. Among the sampled designs, deepening the diamond dimple type appears beneficial ($0.65\%$ per 10 wall units), whereas deepening the triangle types is catastrophic ($-3.3\%$ or $-7.8\%$ per 10 wall units). However, since the steep triangle-up slope rests on a small sample ($n=12$), its magnitude is to be regarded as indicative. The diamond--triangle contrast is nonetheless clear. The diamond dimple type thus exemplifies the classical dimple paradox: while a surface indentation would be expected to add drag, its cavity instead reduces turbulent drag by reshaping the near-wall boundary layer flow rather than simply roughening the surface.

\begin{table}[t]
\caption{Within-type sensitivity of $\mathrm{DR}$ to depth (\% per 10 wall units)}
\label{tab:depthslope}
\centering
\begin{tabularx}{\linewidth}{c >{\raggedright\arraybackslash}X c}
\hline\hline
$\bm{\tau}$ & \textbf{Dimple Type} & $\bm{\Delta(\mathrm{DR})/\Delta d^+}$ \\
\hline
\parbox[c][.75cm][c]{1em}{$\mathbbold{4}$} & Diamond & $0.65$ \\
\parbox[c][.75cm][c]{1em}{$\mathbbold{2}$} & Teardrop-down & $0.03$ \\
\parbox[c][.75cm][c]{1em}{$\mathbbold{3}$} & Teardrop-up & $-0.33$ \\
\parbox[c][.75cm][c]{1em}{$\mathbbold{0}$} & Tapered cylinder & $-0.39$ \\
\parbox[c][.75cm][c]{1em}{$\mathbbold{1}$} & Spherical & $-1.64$ \\
\parbox[c][.75cm][c]{1em}{$\mathbbold{5}$} & Triangle-down & $-3.26$ \\
\parbox[c][.75cm][c]{1em}{$\mathbbold{6}$} & Triangle-up & $-7.76$ \\
\hline\hline
\end{tabularx}
\end{table}

\begin{figure}[t]
\centering
\includegraphics[width=0.9\textwidth]{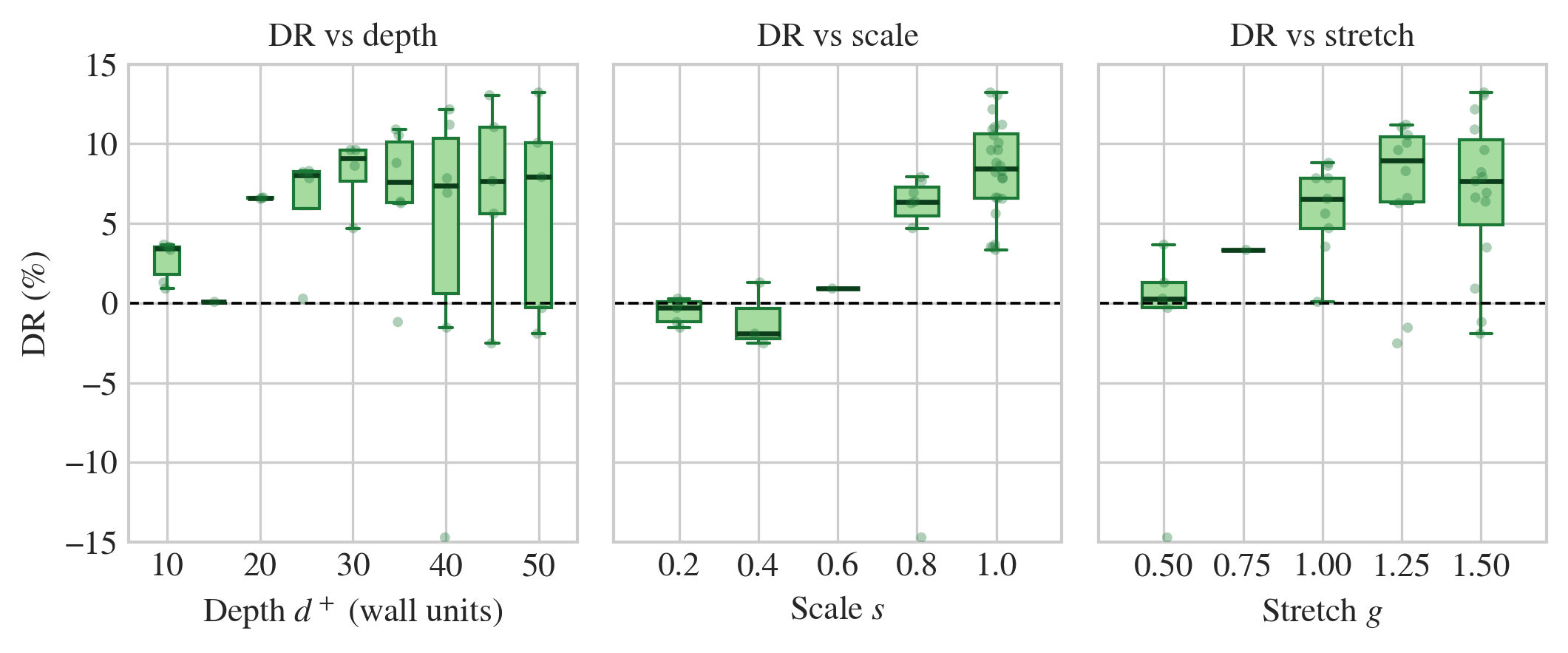}
\caption{Within-diamond ($\tau = \mathbbold{4}$) $\mathrm{DR}$ versus depth, scale, and stretch: per-level
box plots (median and inter-quartile range) and the evaluated designs overlaid
(points).}
\label{fig:diamond}
\end{figure}

Within the diamond dimple design samples, the robustly favorable type in our design space with dense samples ($n=40$), we read its design-parameter trends (see Figure~\ref{fig:diamond}). Scale $s$, or dimple coverage, shows a strong positive influence: $\mathrm{DR}$ rises steeply with $s$, from near-neutral at the partial coverages to strongly drag-reducing at the full coverage ($s=1.0$). Stretch $g$, or streamwise elongation, acts in the same direction; $\mathrm{DR}$ increases with respect to $g$ before plateauing over $g\in[1.25,~1.5]$. On the other hand, the compressed case ($g=0.5$) is the least favorable and hovers near drag-neutral. Depth $d^+$ exerts a gentle positive influence within the sampled diamond dimples, whose aggregate linear trend follows the favorable response to depth noted above (see Table~\ref{tab:depthslope}).

However, it is worth noting that even this favorably drag-reducing dimple type resists a plain single-parameter reading. Although the median $\mathrm{DR}$ increases with these geometric parameters, their inter-quartile spread tends to widen as well. For instance, the deeper diamond dimple surfaces fan out over an increasingly broad range of outcomes rather than settling toward a single value. The coexistence of a positive central slope with growing dispersion is a signature of the complex geometry--flow interaction. The topographic design landscape of dimples is high-dimensional and nonlinear, so the linear sample trends tell only part of the story.

\subsection{Near-Wall Flow Dynamics}\label{sec3.3}

\begin{figure}[t]
\centering
\includegraphics[width=\textwidth]{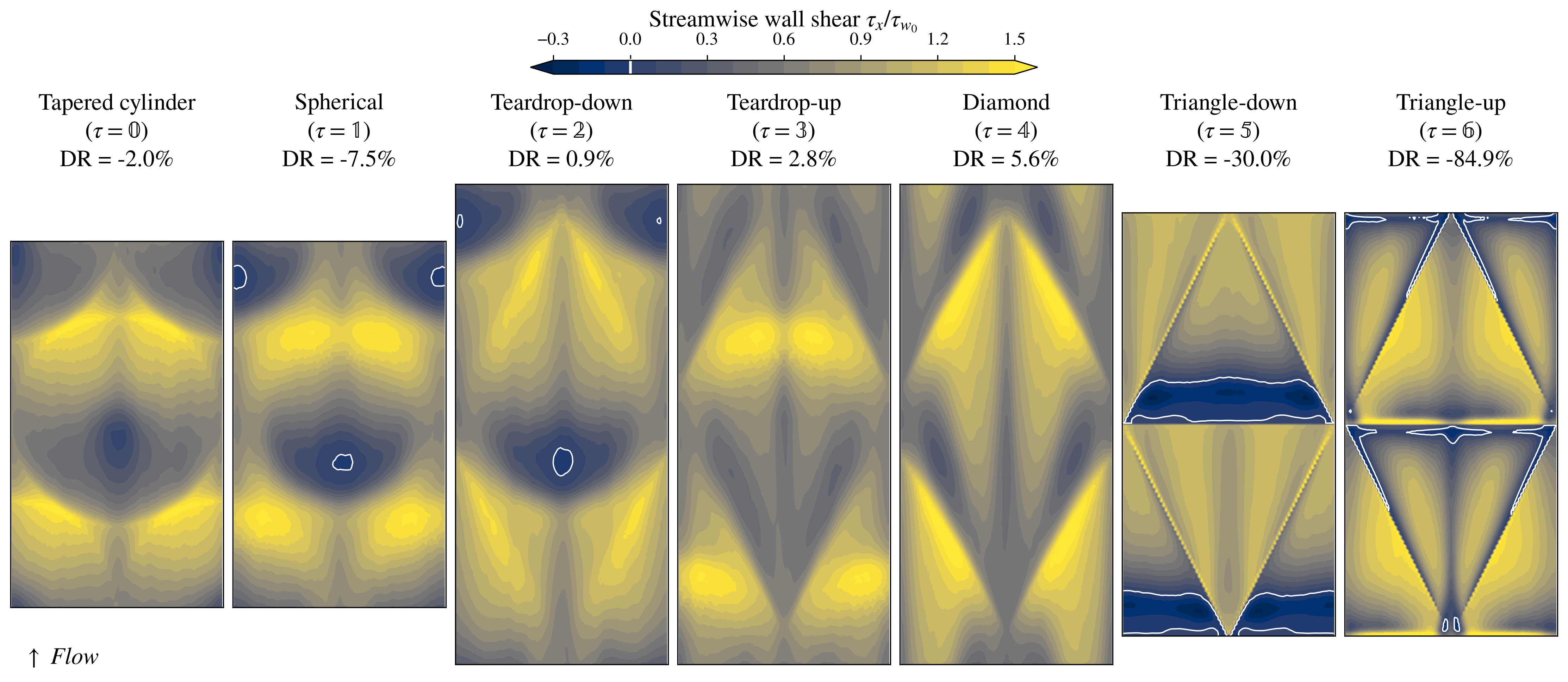}
\caption{Mean streamwise wall shear stress $\tau_x$ on the dimple surface for all
seven types at $(d^+,s,g)=(45,1.0,1.0)$, normalized by the flat channel wall shear
stress $\tau_{w_0}$. Negative $\tau_x$ indicates local flow reversal, depicted by the
$\tau_x=0$ contour line (white).}
\label{fig:flowsections}
\end{figure}

The present LES outcomes, substantially backed by DNS comparisons, resolve the near-wall flow that sets each dimple design's performance. It is therefore worthwhile to read the flow mechanics behind the design ranking. We first compare all seven dimple types at the single matched configuration $(d^+,s,g)=(45,1.0,1.0)$ so that only the topology label differs. On each dimple surface, the streamwise component of the mean wall shear stress $\tau_x$, calculated from the wall-normal velocity gradient with the accurate surface normal extracted from the STL tessellations, is illustrated on the projected $xz-$plane in Figure~\ref{fig:flowsections}. A negative wall shear ($\tau_x<0$) indicates a zone of mean near-wall flow reversal that adds a form drag penalty due to separation. The broad trend is that the better-performing dimple types keep the surface attached, whereas the worse-performing ones develop reversed zones, which leads to the conclusion that suppressing local separation is favorable for drag reduction.

Two comparable contrasts expose the complex geometric levers behind this. First, for a circular planform, the spherical cap ($\tau=\mathbbold{1}$) sustains a compact central reversal pocket that the tapered cylinder ($\tau=\mathbbold{0}$) avoids ($-7.51\%$ versus $-1.87\%$); here the depth profile determines the occurrence of flow separation. Second, changing the orientation of the teardrop dimple switches the $\tau_x < 0$ region on and off ($\tau=\mathbbold{2}$ at $\mathrm{DR}=0.94\%$ has separation, whereas $\tau=\mathbbold{3}$ at $\mathrm{DR}=2.76\%$ keeps the flow attached); in this configuration, the dimple orientation plays a key role in near-wall flow modulation.

The triangle indentations, though the worst performers, are worthy of note. Their non-smooth profiles (a sharp apex and a sloped floor) present the flow with an abrupt step of opposite sense in the two orientations, resembling a backward-facing step ($\tau=\mathbbold{5}$) and a forward-facing step ($\tau=\mathbbold{6}$), respectively; each sheds the extensive reversed-shear region, as shown in Figure~\ref{fig:flowsections}, and the attendant large form drag. Such non-smooth, sharp-peaked topographies are therefore ill-suited to drag-reducing dimple design.

\begin{figure}[t]
\centering
\includegraphics[width=.8\linewidth]{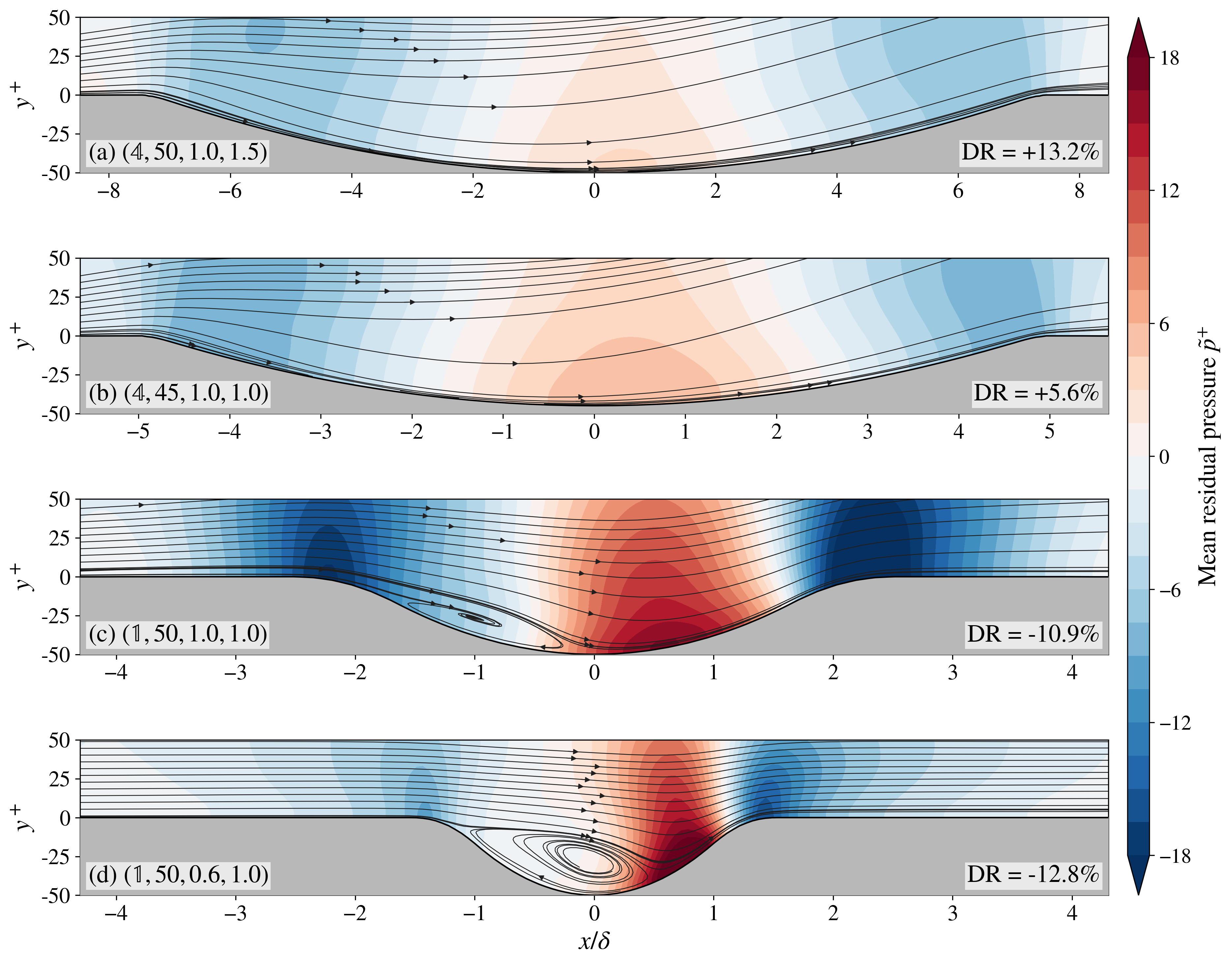}
\caption{Spanwise-centerline ($z=0$) mean residual pressure $\tilde{p}^+$ with
in-plane streamlines, from the dimple bottom up to $y^+=50$, for four representative
designs labeled by $(\tau,d^+,s,g)$: (a) $(\mathbbold{4},50,1.0,1.5)$, (b) $(\mathbbold{4},45,1.0,1.0)$, (c) $(\mathbbold{1},50,1.0,1.0)$, and (d)  $(\mathbbold{1},50,0.6,1.0)$.}
\label{fig:centerline}
\end{figure}

To connect this surface-shear picture to the surrounding flow, Figure~\ref{fig:centerline} contrasts four representative designs on the spanwise-centerline ($z=0$) plane, contouring the mean residual pressure (the periodic part $\tilde{p}$ of Eq.~\eqref{eq:p-decomp}) with the in-plane streamlines from the dimple bottom up to $y^+=50$; each consecutive pair differs chiefly in one factor: streamwise stretch from (a) to (b), topology from (b) to (c), and coverage from (c) to (d). The diamond keeps the near-wall flow attached whether it is streamwise-elongated or not. Nonetheless, the maximum $\tilde{p}$, located slightly past the dimple bottom and inducing form drag, becomes stronger with decreasing stretch. This arises because less stretch increases the concavity's curvature; the surface-attached streamlines then deflect more, raising the local pressure. Switching to the spherical topology, and then thinning its coverage of the surface, instead makes the concavity trap a growing recirculation bubble, seen in the closed near-wall streamlines and the intensifying adverse pressure on its lee side. This is consistent with the DNS of~\citet{Ng2020}, in which the diamond reduces drag while the spherical dimple increases it: less near-wall flow separation goes with less drag.

\subsection{Metamodel Sensitivity Analysis}\label{sec3.4}

Throughout the sample-based dimple design analysis, we discovered that the design factors do not act independently. To explore this nonlinear design landscape beyond the samples, we augment the $150$ design evaluations collected via MixMOBO~\citep{Sheikh2022a} using a Gaussian process (GP) metamodel and a variance-based (Sobol) sensitivity analysis~\citep{Marrel2009,LeGratiet2014,Wirthl2023}. This analysis resolves the cross-parameter interactions that the raw sample trends cannot expose. Such a decomposition is only as trustworthy as the surrogate beneath it; we first establish the reliable metamodel.

The surrogate is a joint GP built upon an anisotropic Mat\'ern-$5/2$ kernel over the continuous design variable $d^+$, $s$ and $g$ combined with a categorical kernel over the categorical variable $\tau$, and is fit to the log-transformed objective $\log|G_m|$ from the evaluated 150 credible LES samples, whose predictions we back-transform to the original objective $|G_m|$ for all reported $Q^2$ and RMSE values. Evaluating this metamodel on every design point of the 1575 dimple candidates, we obtain first-, total- and second-order Sobol\textquotesingle~indices~\citep{Sobol2001,Saltelli2010}. The GP uncertainty is propagated through $N_{\mathrm{GP}}=500$ posterior realizations~\citep{Wirthl2023}.

\begin{figure}[t]
\centering
\includegraphics[width=0.6\linewidth]{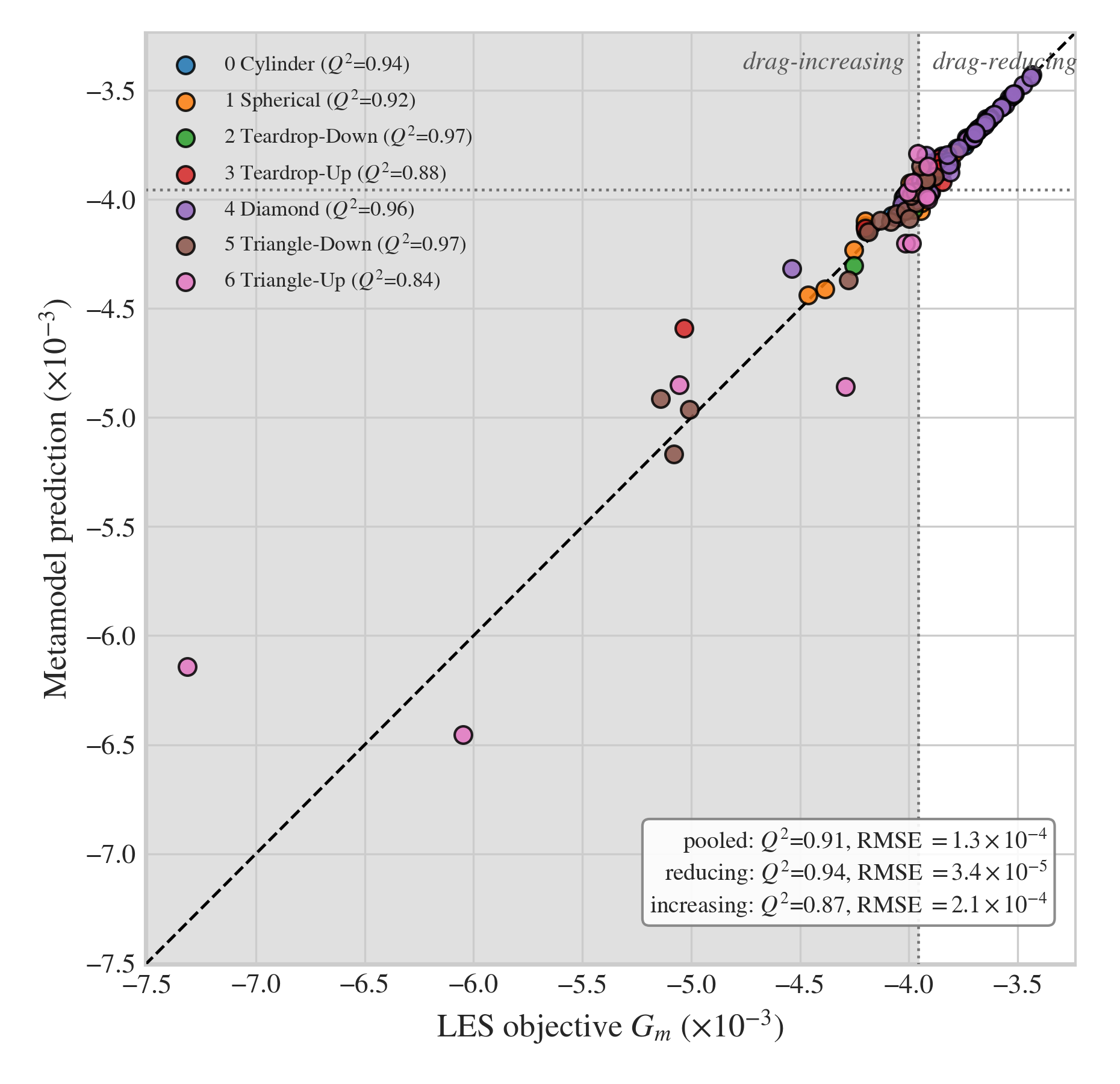}
\caption{Leave-one-out cross-validation results of the GP metamodel.}
\label{fig:loo}
\end{figure}

Owing to the relatively small evaluated sample size of 150, leave-one-out cross-validation (LOOCV) is a feasible choice for validation. As shown in Figure~\ref{fig:loo}, the validation confirms that the metamodel predicts held-out designs well, with a pooled $Q^2=0.91$ (RMSE~$=1.3\times10^{-4}$) and every family well-predicted ($Q^2\ge0.84$). Crucially, in the drag-reducing region ($G_m > G_m^{\mathrm{flat}} = -3.956 \times 10^{-3}$) the metamodel reaches $Q^2 = 0.94$ (RMSE~$=3.4\times10^{-5}$), which indicates a strong predictability, while degrading to $Q^2 = 0.87$ (RMSE~$=2.1\times10^{-4}$) only in the drag-increasing side. This is clearly due to the BO run's intended bias; the Bayesian optimizer deliberately sought drag-reducing dimple designs, and the evaluated designs cluster densely in the drag-reducing region. Therefore, the surrogate is best resolved precisely where the design insights of interest reside.

Table~\ref{tab:sobol1} reports the first- and total-order Sobol\textquotesingle~indices $S_i$ and $S_{Ti}$. Dimple type is clearly the most influential design factor ($S_i=0.289$, $S_{Ti}=0.681$). The remaining design factors, however, tell a subtler story: scale carries a small first-order index ($S_i=0.065$) yet a total-order index some seven times larger ($S_{Ti}=0.492$). In other words, it governs drag reduction almost entirely through interaction rather than on its own. Stretch ($S_i=0.035$, $S_{Ti}=0.129$) and depth ($S_i=0.102$, $S_{Ti}=0.272$) follow the same interaction-mediated pattern at lower amplitude. Thus, it can be concluded that no continuous parameter considered in the present study acts independently of dimple type.

\begin{table}[t]
\caption{First-order ($S_i$) and total-order ($S_{Ti}$) Sobol\textquotesingle~indices with
metamodel uncertainty.}
\label{tab:sobol1}
\centering
\begin{tabularx}{\linewidth}{>{\raggedright\arraybackslash}X c c c}
\hline\hline
\textbf{Factor} & $\bm{S_i}$ & $\bm{S_{Ti}}$ & $\bm{S_{Ti}-S_i}$ \\
\hline
Dimple type ($\tau$) & $0.289\pm0.004$ & $0.681\pm0.005$ & 0.392 \\
Depth ($d^+$) & $0.102\pm0.002$ & $0.272\pm0.007$ & 0.170 \\
Scale ($s$) & $0.065\pm0.001$ & $0.492\pm0.004$ & 0.427 \\
Stretch ($g$) & $0.035\pm0.001$ & $0.129\pm0.003$ & 0.094 \\
\hline\hline
\end{tabularx}
\end{table}

The second-order Sobol\textquotesingle~indices (see Table~\ref{tab:sobol2}) demonstrate where the pairwise interactions live: the dominant interaction is type $\times$ scale ($S_{ij}=0.296$), roughly six times the next-largest coupling (type $\times$ depth, $0.050$), while the purely continuous--continuous interactions each stay small (less than or equal to $0.035$). Dimple type thus gates the interaction structure; once it is fixed, the continuous design factors couple weakly among themselves, yet none of them can be assessed independently and apart from the chosen dimple type. First- and second-order effects together account for $93.6\%$ of the output variance, meaning that the response is well-captured by these main effects plus pairwise couplings.

\begin{table}[t]
\caption{Second-order Sobol\textquotesingle~indices $S_{ij}$ with metamodel uncertainty.}
\label{tab:sobol2}
\centering
\begin{tabularx}{\linewidth}{>{\raggedright\arraybackslash}X c}
\hline\hline
\textbf{Interaction pair} & $\bm{S_{ij}}$ \\
\hline
Dimple type $\times$ Scale & $0.296\pm0.005$ \\
Dimple type $\times$ Depth & $0.050\pm0.002$ \\
Scale $\times$ Stretch & $0.035\pm0.001$ \\
Depth $\times$ Scale & $0.034\pm0.001$ \\
Depth $\times$ Stretch & $0.028\pm0.001$ \\
Dimple type $\times$ Stretch & $0.002\pm0.000$ \\
\hline\hline
\end{tabularx}
\end{table}

\begin{figure}[t]
\centering
\includegraphics[width=0.7\textwidth]{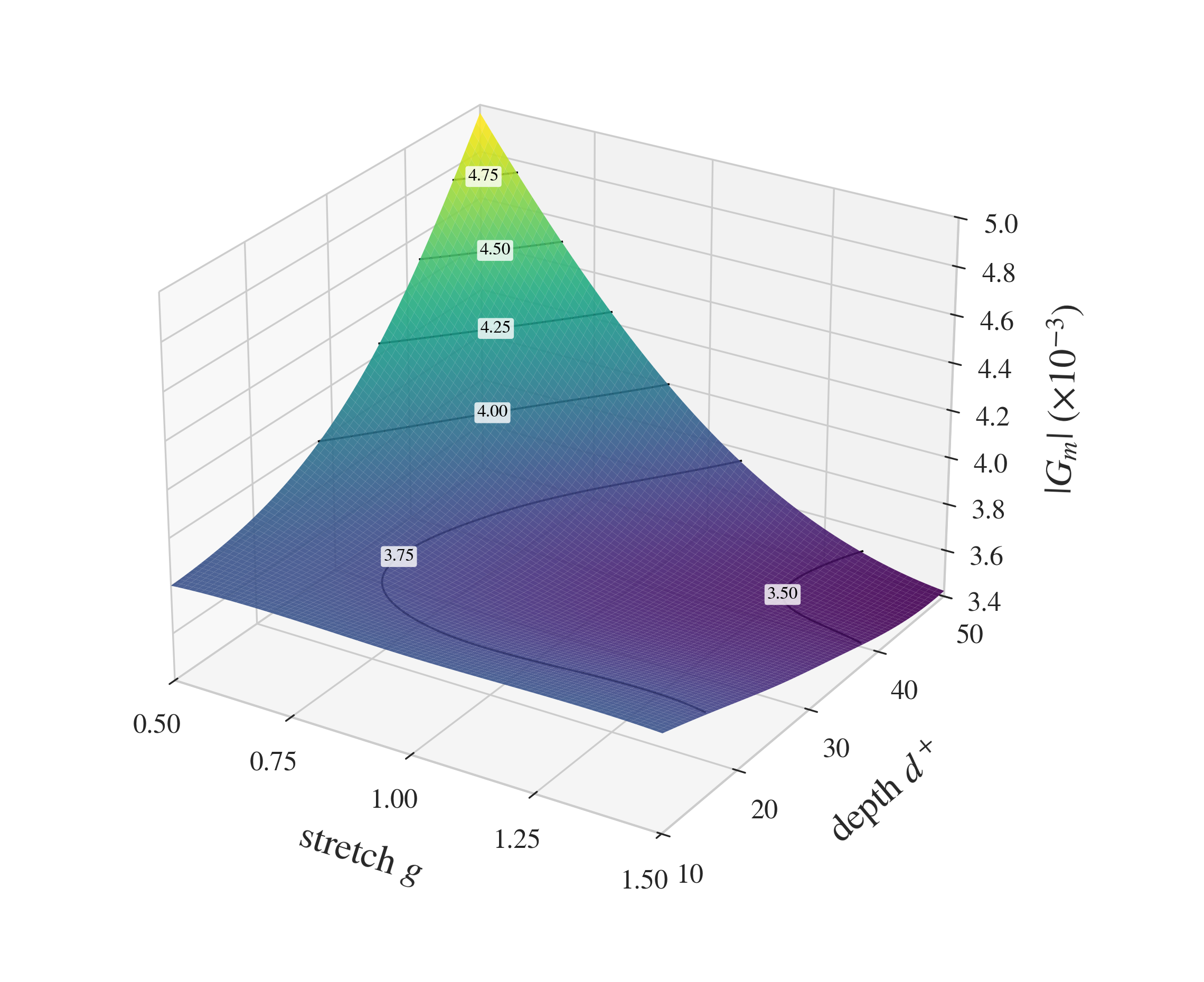}
\caption{GP metamodel response surface for the diamond at full coverage
($\tau=\mathbbold{4}$, $s=1.0$): $|G_m|$ ($\times10^{-3}$) over depth $d^+$ and
stretch $g$, with isolevels drawn on the surface.}
\label{fig:rs}
\end{figure}

These interactions overturn a conclusion that the sample-based observation would support. In Table~\ref{tab:depthslope} and Figure~\ref{fig:diamond}, the aggregated factor-by-factor trends make deepening the diamond concavities appear beneficial. However, on the metamodel that benefit is found to be confined to the full-coverage, streamwise-elongated design region. Figure~\ref{fig:rs} presents the GP metamodel's $|G_m|$ response over depth $d^+$ and stretch $g$ for the diamond ($\tau=\mathbbold{4}$) at full coverage ($s=1.0$). Deepening acts favorably in concert with streamwise elongation (e.g., at $g=1.5$). Across the rest of the drag-reducing region, the depth response flattens and even turns adverse when we consider streamwise-contracted dimpled surfaces (e.g., $|G_m|$ increases with respect to $d^+$ at $g=0.5$). The conclusion is clear: neither deeper nor shallower depth is unconditionally beneficial for drag reduction.

This behavior reaffirms that the dimple design landscape is nonlinear and interaction-mediated. A classical factor-by-factor analysis misses conditional influences of design factors due to coupling, revealing only part of the story and risking misleading conclusions. Instead, the metamodel sensitivity analysis provides an augmented exploration of the design space. The GP interpolates and extends the drag-reducing-focused BO evaluations to map a performance landscape that direct sampling alone cannot, and the surrogate's improved resolution exposes the important couplings: topology-gated and interaction-dominated.

\subsection{Empirical Design Insights}
\label{sec3.5}

Drawing together the Bayesian optimization run and sensitivity results within the dimple design space, we suggest the following conditional dimpled surface design guidance for drag reduction in turbulent channel flow corresponding to the $Re_\tau = 180$ flat channel flow rate condition:
\begin{enumerate}
    \item[(i)] \textit{Select the dimple topology first.} The dimple shape is by far the most influential factor: only the diamond type reaches substantial drag reduction robustly, whereas the others remain neutral or strongly increase drag. The brief fluid-mechanistic analysis reveals that the better dimples are those that keep the flow attached without reversal or separation, largely owing to their moderate curvature.
    \item[(ii)] \textit{Cover the surface fully.} Coverage matters in combination with a favorable topology, and drag reduction can grow as the dimples are packed more densely; at partial coverage the exposed flat wall between cavities dilutes the influence of introducing dimples.
    \item[(iii)] \textit{Elongate the dimple in streamwise direction.} Streamwise elongation can aid drag reduction, as seen both in the present LES evaluations and in the DNS of~\citet{Ng2020} (spherical versus elliptical dimples), whose elongated dimple outperforms its counterpart. 
    \item[(iv)] \textit{Set the depth empirically.} Depth carries no universal sign within the explored design space. It is best chosen directly once the preceding choices have fixed the favorable design region.
\end{enumerate}
We do not, however, claim that this guidance is exclusive; rather, we offer it as one plausible pathway toward achieving drag-reducing dimples. Because the design space is governed by intricate cross-factor interactions, we assert that no single, universal parametric prescription for drag-reducing dimples exists. In the context of designing dimpled surfaces, the data-driven design exploration and sensitivity analysis adopted here can be preferred over a classical factor-by-factor design study.

\section{Discussion}\label{sec4}

First of all, we draw attention to the primacy of the dimple topology. Across the design space, the topology of a dimple unit is by far the most influential factor, and robustly favorable drag reduction is confined to a single dimple type: the diamond. This optimization result is substantial: the diamond was already the best baseline in the DNS of~\citet{Ng2020}, yet tuning the depth, scale, and stretch variables raises its performance to $13.2\%$. This value is well above the few-percent reductions typically reported for turbulent-flow dimples~\citep{Tay2015,vannesselrooij2016}.

Now that topology is known to be the key factor for dimple drag reduction, the categorical choice among a fixed set of dimple types stands out as a natural limitation to move beyond. Rather than selecting from a limited catalog of dimple types, one could treat the topology itself as a continuously variable design object through classical topology optimization or Design-by-Morphing (DbM)~\citep{Lee2024,Sheikh2023,Lee2026} for the exploration of more dimple topographies.

Next, we consider why the scale and elongation of the optimal design improve its performance. A similar trend appears in the DNS of~\citet{Ng2020}: comparing the spherical dimple with its stretched (elliptical) version, the drag-reduction performance flips sign, from drag-increasing to drag-reducing. This appears consistent with a competition intrinsic to wall textures: a streamwise channeling of the near-wall flow, which lowers drag, set against a separation-driven form-drag penalty~\citep{Tay2015,Ng2020}. Seen this way, the optimum is less an array of discrete concavities than an emergent groove-like texture. At full coverage the elongated cavities resemble a continuous, nearly streamwise-consistent groove network, presumably acquiring the defining, drag-reducing feature of riblets~\citep{Garcia-Mayoral2011}. We postulate that this shared morphology is a plausible connection between the optimized dimple and riblets.

The Bayesian optimizer's tendency to favor drag-reducing designs during sampling should also be noted. The optimizer deliberately concentrates its evaluations in promising regions, so the collected design samples over-represent drag-reducing topographies. This bias is, however, intentional: our aim was to characterize which dimples reduce drag, not to estimate how common drag reduction is across the space. The GP metamodel inherits this bias and, rather than correcting it, augments the sparse gaps between the few affordable large eddy simulations. Its resolution is therefore sharpest where the sampling was densest, in the drag-reducing region where the design insights live. The proposed design insights are correspondingly less applicable in the sparsely explored drag-increasing region, which is a deliberate cost of concentrating the search on the best design objective.

Last but not least, two limitations bound the discovered optimum. The first is that it lies at the very boundary of the explored ranges: the deepest, most streamwise-elongated diamond considered. Extrapolating the response trend beyond the present box (see Figure~\ref{fig:rs}), one may expect still-deeper or more elongated designs to perform better, though only marginally; a refined search over a broadened space is a natural next step. The second is that every design is evaluated at a single friction Reynolds number, $Re_\tau=180$, and tests at the higher Reynolds numbers of practical flows will be needed. It is noted that the boundary layer measurements of~\citet{vannesselrooij2016} reported a drag reduction that improved with increasing Reynolds number, suggesting that the present $Re_\tau$ setting adopted here is conservative. We therefore expect the acquired design insights here (topology first, with coverage and elongation acting through interactions) to carry across Reynolds number more robustly than the absolute rates, since the design variables are specified in viscous wall units and a higher Reynolds number therefore rescales the same design rather than posing a different flow problem. As the Reynolds number rises, however, that same scaling shrinks the dimple physically, and durability and contamination may become the governing constraints~\citep{spalart2011industry}.

\section{Conclusion}\label{sec5}

We conducted a 150-evaluation Bayesian optimization run over a four-parameter (dimple type $\tau$, depth $d^+$, scale $s$, and stretch $g$) dimple design space with validated large eddy simulations for turbulent channel flow at the constant flow rate corresponding to the $Re_\tau=180$ flat channel case. The identified optimum was at $\tau=\mathbbold{4}$ (diamond; see Figure~\ref{fig:baseline_stl}), $d^+=50$, $s=1.0$, and $g=1.5$ (see Figure~\ref{fig:surface_schema}), giving a $13.2\%$ drag reduction relative to the flat channel wall; the drag-reduction rate is defined in Eq.~\eqref{eq:DR}. Our findings confirm that dimpled texturing is only conditionally beneficial: among the $150$ evaluated designs, $63\%$ reduce drag and $37\%$ increase it, spanning $13.2\%$ to $-84.9\%$. Because these designs were selected by an optimizer that preferentially samples promising regions, this proportion is likely to overstate the prevalence of drag reduction across the design space. Nonetheless, this division clearly demonstrates that neither drag reduction nor drag increase is a universal property of dimpled surfaces.

A Sobol\textquotesingle~sensitivity analysis on the augmented Gaussian process metamodel established topology as the master variable (first-order Sobol\textquotesingle~$S_i=0.29$, total-order $S_{Ti}=0.68$), with the diamond being the only type that reduces drag robustly across the design space. Scale and stretch, by contrast, act as conditioning levers rather than main effects---each shows a small $S_i$ but a disproportionately larger $S_{Ti}$ ($0.49$ and $0.13$, respectively)---with the topology$\times$scale interaction dominating ($S_{ij}=0.30$). Depth is a secondary, interaction-mediated factor that carries no universal sign: it turns favorable to the diamond only within the full-coverage, streamwise-elongated corner where the optimum sits, which is why its aggregate correlation is uninformative.

A few promising directions remain for future study. As the discovered optimum lies at the boundary of the explored ranges, a broader design space with a refined search can yield further gains; the friction Reynolds number itself, fixed here at $Re_\tau=180$ for its rich direct numerical simulation database, could be promoted to an additional design input, which the present pipeline already accommodates. The dimple type, currently a discrete and categorical choice, could likewise be treated as a continuous design variable (e.g., by topological morphing or material removal and addition). Above all, the underlying physical mechanisms linking dimple topography to turbulent drag reduction (only partially inferred here) call for a dedicated fluid physics study that contrasts the best and worst designs through their near-wall flow and vorticity fields, clarifying why some dimple topographies decrease drag while others amplify it.

\section*{CRediT authorship contribution statement}
\textbf{S.L.:} Conceptualization, Software, Data Curation, Formal Analysis, Investigation, Methodology, Writing – Original Draft, Writing – Review \& Editing. \textbf{M.E.Y.:} Investigation, Methodology, Writing – Review \& Editing. \textbf{H.M.S.:} Conceptualization, Software, Project Administration, Resources, Supervision, Writing – Review \& Editing.

\section*{Acknowledgements}
This work used Anvil at Purdue RCAC through allocation PHY250071 from the Advanced Cyberinfrastructure Coordination Ecosystem: Services \& Support (ACCESS) program, which is supported by U.S. National Science Foundation grants \#2138259, \#2138286, \#2138307, \#2137603, and \#2138296.


\bibliographystyle{elsarticle-num-names}
\bibliography{bibliography}

\end{document}